\documentstyle[aaspp4,psfig,11pt]{article}

\slugcomment{\it Submitted to ApJ}

\begin{document}

\title{Weak Lensing Measurements: A Revisited Method \\
and Application to HST Images}

\author{Jason Rhodes}
\affil{Physics Department, Princeton University, Jadwin Hall,
P.O. Box 708, Princeton, NJ 08544; jrhodes@pupgg.princeton.edu}
\and
\author{Alexandre Refregier\footnote{present address: Institute
of Astronomy, Madingley Road, Cambridge, CB3 OHA, U.K.,\\ email:
ar@ast.cam.ac.uk}}
\affil{Department of Astrophysical Sciences, Princeton University,
Princeton, NJ 08544; refreg@astro.princeton.edu}
\and
\author{Edward J. Groth}
\affil{Physics Department, Princeton University, Jadwin Hall,
P.O. Box 708, Princeton, NJ 08544; groth@pupgg.princeton.edu}

\begin{abstract}
The weak distortions produced by gravitational lensing in the images
of background galaxies provide a unique method to measure directly the
distribution of mass in the universe. However, because the induced
distortions are only of a few percent, this technique requires high
precision measurements of the lensing shear and cautious corrections
for systematic effects. Kaiser, Squires, \& Broadhurst (1995) proposed
a method to calibrate the ellipticity--shear relation in the presence
of Point Spread Function (PSF) anisotropies and camera
distortions. Here, we revisit the KSB method in the context of the
demanding search for weak lensing by large-scale structure. We show
that both the PSF and the camera distortions can be corrected for
using source moments, as opposed to ellipticities. We clarify the
applicability of some of the approximations made in this method.  We
derive expressions for the corrections which only involve the galaxy
moments. By decomposing the moments into spinors, we derive an
explicit relation between the shear and the average ellipticity. We
discuss the shortcomings of the method, and test its validity using
numerical simulations. As an application of the method, we repeat the
analysis of the HST--WFPC2 camera performed by Hoekstra et
al. (1998). We confirm the presence of sizable ($\sim 10\%$) PSF
ellipticities at the edge of the WFPC2 chips. However, we find that
the camera distortion is radial, rather than tangential. We also show
that the PSF ellipticity varies by as much as 2\% over time. We use
these measurements to correct the shape of galaxies in the HST Survey
Strip (``Groth'' Strip). By considering the dependence of the
ellipticities on object size, we show that, after corrections, the
residual systematic uncertainty for galaxies with radii greater than
0.15 arcsec, is about 0.4\%, when averaged over each chip. We discuss
how these results provide good prospects for measuring weak lensing by
large-scale structure with deep HST surveys.
\end{abstract}

\section{Introduction}
Weak gravitational lensing produces coherent distortions in the images
of background galaxies. This effect provides a unique method to
measure directly the distribution of mass in the universe (for reviews
see \markcite{sch92}Schneider et al. 1992; \markcite{nar96}Narayan \&
Bartelmann 1996; \markcite{mel99}Mellier 1999).  This technique is now
routinely used to map the mass of clusters of galaxies (for a review,
see Fort \& Mellier 1994\markcite{for94}). A search for weak lensing
by large-scale structure is the subject of much recent and on-going
theoretical and observational effort (eg. \markcite{vil95}Villumsen
1995; \markcite{ste95}Stebbins, et al. 1995; \markcite{kai96}Kaiser
1996; \markcite{sch97}Schneider et al. 1997; \markcite{van98}Van
Waerbeke et al. 1998; \markcite{refe98}Refregier et al. 1998; see
\markcite{ref99}Refregier 1999 for a bibliography). The main
difficulty lies in the fact that the lensing distortions are small
($\sim 10\%$ for clusters and $\sim 1\%$ for large-scale structure),
thus requiring high precision measurements and tight control of
systematic effects.

\markcite{kai95}Kaiser, Squires and Broadhurst (1995, KSB) have
developed a method to correct for the major systematic effects, namely
the anisotropy of the Point-Spread Function (PSF) and camera
distortions, and to calibrate the relation between galaxy
ellipticities and lensing shear (for other methods, see also
\markcite{bon95}Bonnet \& Mellier 1995; \markcite{sch95}Schneider \&
Seitz 1995). Further elements of their method were presented in
\markcite{lup97}Luppino \& Kaiser (1997, LK) and in
\markcite{hoe98}Hoekstra et al. (1998, HFKS). 

Recently, \markcite{kai99}Kaiser (1999) pointed out that the KSB
method had several shortcomings, all stemming from the fact that most
PSFs encountered in practice are not sufficiently compact.  Kaiser
then proposed another method based on the explicit construction of the
post-convolution shear operator. Another alternative method was
recently proposed by \markcite{kui99}Kuijken (1999). In this different
approach, a sheared and convolved isotropic model is fitted to the
galaxy image, so as to derive an estimator for the shear. These two
methods are promising, but both require complete knowledge of the
2-dimensional PSF function, while PSF measurements are sparse due to
the finite number of stellar images.

Here, we revisit the KSB method, which has the advantage both of being
linear and of relying only on the first multipole moments of the PSF
and galaxy images.  We focus on the demanding search for weak lensing
by large-scale structure. We show that both the PSF and the camera
distortions can be corrected for using source moments, as opposed to
ellipticities. We clarify the applicability of some of the
approximations made in this method, and show how the weight function
for stars can be chosen to be different from that for galaxies. We
derive expressions for the corrections in term of the moments only. By
decomposing the moments into spinors, we derive an explicit relation
between the shear and the average ellipticity. We discuss the
shortcomings of the method discussed in \markcite{kai99}Kaiser (1999)
and \markcite{kui99}Kuijken (1999), and test its validity using
numerical simulations.

As an application, we consider weak lensing measurements with the
Hubble Space Telescope (HST). The small PSF and absence of atmospheric
seeing makes HST an ideal instrument for weak lensing measurements
(eg. \markcite{kne96}Kneib et al. 1996; HFKS). We reproduce the
analysis of HFKS who studied, in detail, the PSF and camera distortion
of the WFPC2 camera onboard HST. We apply these calibrations to the
galaxies in the HST Survey Strip known as the ``Groth Strip'' (Groth et
al. 1995\markcite{gro95}, \markcite{rho99}Rhodes 1999). In particular,
we show how the PSF anisotropy and camera distortions depend on the
size of the galaxies. We also discuss the prospects of weak lensing
measurements with HST. A description of our search for weak lensing by
large-scale structure with the Survey Strip will be presented in
Rhodes, Refregier, \& Groth 1999\markcite{rho99} (see also
\markcite{rhot99}Rhodes 1999).

Because the method is somewhat complex, we provide a practical summary
in \S\ref{overview} in which we point to the results and equations
which are of direct practical interest. In \S\ref{shapes}, we describe
measures of object shapes. Next, in \S\ref{distortion} and
\S\ref{convolution}, we show how these measures are affected by the
two main classes of deformations, namely convolutions and distortions,
and derive explicit expressions to correct for them.  In \S\ref{comb},
we study the effect of shear combined with a convolution on the
observed ellipticity. In \S\ref{validity}, we discuss the shortcomings
of the method, and test its validity using numerical simulations. In
\S\ref{HST}, we apply our method to HST observations. In particular,
we consider measurements of the camera distortion, globular cluster
observations, and finally apply our results to the Survey Strip.  Our
conclusions are summarized in \S\ref{conclusions}.

\section{Overview of the Method}
\label{overview}
The purpose of this method is to provide a measure of the shapes of
galaxies, to correct for instrumental effects, and to derive an
estimate for the weak lensing shear. The images of galaxies are
assumed to be altered by two kinds of operations: {\it distortions}
(sometimes simply called shear), which are intensity-conserving
mappings between the source plane and the image plane (see
Eq.[\ref{eq:fprime}]), and {\it convolutions} (or smear; see
Eq.[\ref{eq:iprime_conv}]). Specifically, we follow KSB and HFKS and
assume that a galaxy image is altered by a distortion due to
gravitational lensing, followed by a convolution by a generally
anisotropic PSF, and then by a distortion due to the camera
optics. Schematically, we can write
\begin{equation}
{\rm intrinsic~image} \rightarrow {\rm lensing} (\partial)
\rightarrow {\rm PSF} (*) \rightarrow {\rm camera} (\partial)
\rightarrow {\rm observed~image},
\end{equation}
where $\partial$ and $*$ symbolize distortions and convolutions,
respectively. Here, we revisit the KSB method. We show how to correct
for the two latter effects and how to calibrate the sought-after
effect of lensing.

\begin{figure}
\epsscale{0.8}
\plotone{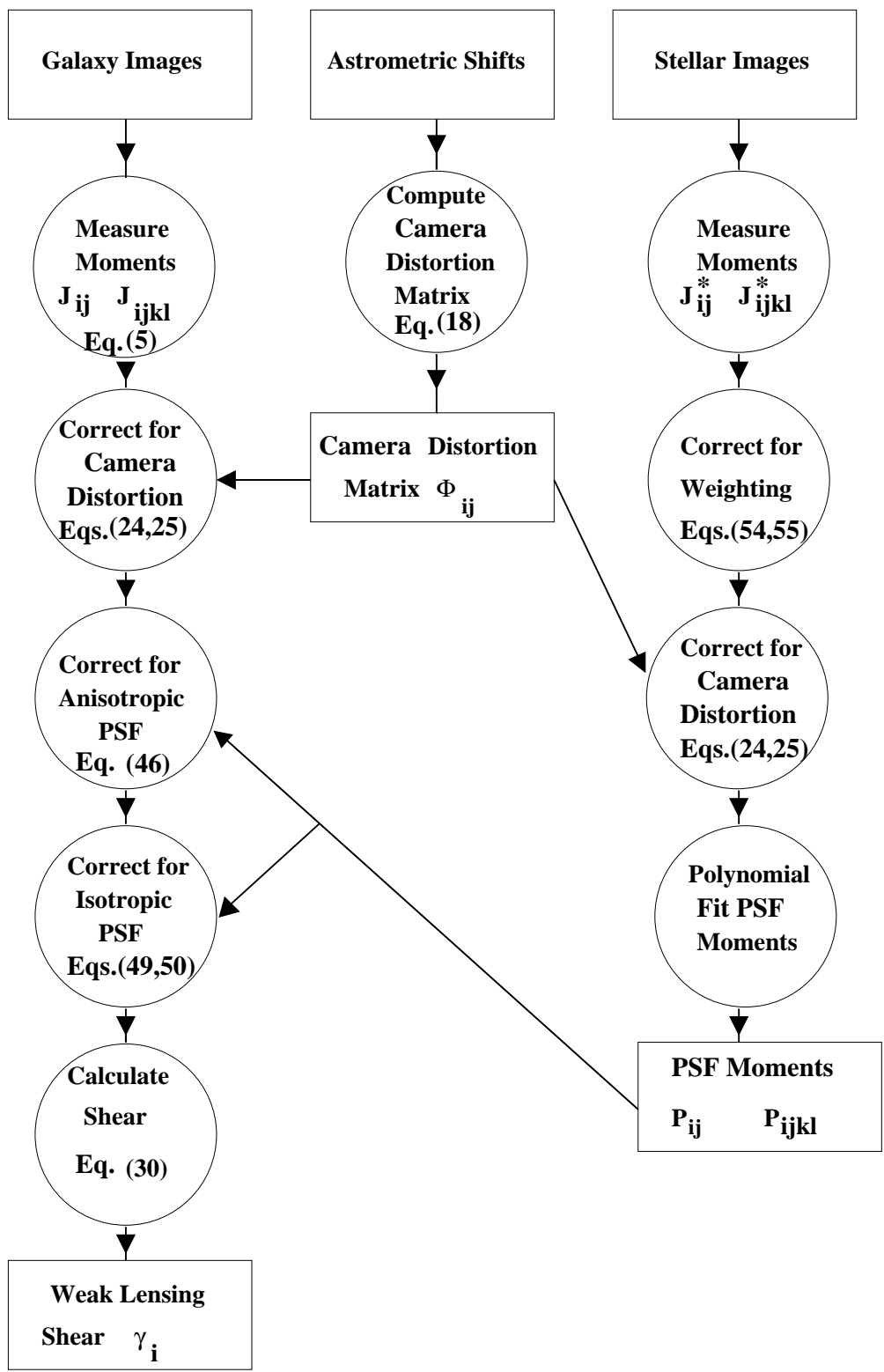}
\caption{Synopsis of the weak lensing measurement method.}
\label{fig:synopsis}
\end{figure}

The synopsis of our method is shown on figure~\ref{fig:synopsis}.
Since the method is somewhat complex, we provide here a summary of the
method, and point to the equations which are of direct practical
interest. References to these equations can also be found on this
figure.

The basis of the method is to characterize the shapes of galaxies and
stars by measuring their multipole moments $J_{ij}$, $J_{ijkl}$, etc.
(Eq.~[\ref{eq:jij_def}]). To enforce convergence in the presence of
image noise, these moments are measured with a weight function
$w(\theta)$ which we choose to be a gaussian of width $\omega$
(Eq.~[\ref{eq:w_Gaussian}]). This choice allows us to write all the
expressions for the corrections in terms of the moments. Unlike KSB,
we perform all the corrections using moments, and postpone the use of
ellipticities (Eq.~[\ref{eq:epsilon_def}]) until the last step. This
has the advantage of keeping track of the size of the PSF
and of the galaxies, and thus of reducing the noise resulting from the
corrections, if the PSF size varies across the field.

The first step consists of deriving the distortion matrix
$\phi_{ij}^{\rm camera}$ (Eq.~[\ref{eq:phi_def}]) for the camera
distortion. HFKS showed that this could be achieved by considering
astrometric shift solutions, such as that of \markcite{hol95}Holtzmann
et al. (1995) for the WFPC2 camera on HST. The resulting shear pattern
for WFPC2 is shown on figure~\ref{fig:wf_shear}. (See discussion
in \S\ref{hfks_camera} about the difference between this figure and
the results of HFKS).

In the second step, we derive the PSF moments from stellar images. For
this purpose, the stellar multipole moments $J_{ij}^{*}$ and
$J_{ijkl}^{*}$ are measured using an optimally chosen weight-function
width $\omega_{*}$. These moments are then corrected for the weight
function to provide (an approximation to) the unweighted moments of
the PSF
(Eqs.~[\ref{eq:pij_correction},\ref{eq:pijkl_correction}]). This
allows us to use a different weight-function width $\omega$ for the
galaxies, and thus to improve the sensitivity. The PSF moments are
then corrected for the camera distortion
(Eqs~[\ref{eq:jij_dcorrection},\ref{eq:jijkl_dcorrection}] with
$\omega \rightarrow \infty$). This provides the corrected unweighted
PSF moments $P_{ij}$ and $P_{ijkl}$
(Eqs.~[\ref{eq:pij_def},\ref{eq:pijkl_def}]), that can be then be
interpolated across the field using low-order polynomial fits for each
component.  Figures~\ref{fig:m4ellipt} and \ref{fig:ngc6572} show the
resulting ellipticities of the WFPC2 PSF derived from two globular
clusters, while figure~\ref{fig:m4sm} shows the interpolated PSF
ellipticities derived from a combination of the two after the moments
have been corrected for weighting and camera distortion.

In the third step, we measure the galaxy moments and correct them for
instrumental effects. The galaxy moments $J_{ij}$ and $J_{ijkl}$ are
measured using an optimal weight-function width $\omega$.  They are
then corrected for the camera distortion using $\phi_{ij}^{\rm
camera}$
(Eqs.~[\ref{eq:jij_dcorrection},\ref{eq:jijkl_dcorrection}]). The PSF
can be decomposed into an anisotropic and an isotropic part
(Eq.~[\ref{eq:p_decomp}]). We correct the galaxy moments $J_{ij}$ for
the PSF anisotropy, by expanding in powers of the PSF ellipticity
(Eq.~[\ref{eq:jij_acorrection}]). Since, by construction, the
isotropic part of the PSF is a gaussian (Eq.~[\ref{eq:pi_Gaussian}]),
we can perform the isotropic correction exactly for $J_{ij}$
(Eq.~[\ref{eq:jij_icorrection}]). The fourth order moments $J_{ijkl}$
are then approximately corrected for the PSF
(Eq.~[\ref{eq:jijkl_correction}]). This provides us with the corrected
galaxy moments $J_{ij}^{\omega_{g}}$ and $J_{ijkl}^{\omega_{g}}$ which
are now effectively weighted by the new weight-function width
$\omega_{g}$ (Eq.~[\ref{eq:omega_g}]).

The final step consists of measuring the weak lensing shear by
averaging over an ensemble of galaxies in a region of the sky. For
this purpose, we compute the ellipticity $\epsilon_{i}$ of each galaxy
from its corrected moments $J_{ij}^{\omega_{g}}$
(Eq.~[\ref{eq:epsilon_def}]). The weak lensing shear
$\gamma_{i}$ is then computed from the average ellipticity $\langle
\epsilon_{i} \rangle$ (Eq.~[\ref{eq:gamma_epsilon}]). In this last
step, one should remember to use $\omega_{g}$ rather than $\omega$ as
the weight-function width.  The relationship between $\gamma_{i}$ and
$\langle \epsilon_{i} \rangle$ is greatly simplified by considering
the rotation properties of the multipole moments (see
\S\ref{rotations}). Note that our method avoids the complications of
the pre-smear and post-smear shear susceptibilities discussed in LK
and HFKS. In \S\ref{HST}, we discuss the application of our method
to the HST Survey Strip.

\section{Source Shape Characterization}
\label{shapes}
In this section, we show how object shapes can be characterised using
multipole moments and related quantities. We also study the rotational
properties of the moments and decompose them into spinor
representations.

\subsection{Moments} 
\label{Moments}
Let us consider a source with intensity $i({\mathbf \theta})$. As a
first step, we find the centroid ${\mathbf \theta}^{0}$ of the source
by solving
\begin{equation}
\label{eq:centroid}
\int d^{2}\theta ~(\theta_{i}-\theta_{i}^{0}) w({\mathbf \theta} -
  {\mathbf \theta}^{0}) i({\mathbf \theta}) = 0,
\end{equation}
where $w(\theta)$ is a weight function introduced to ensure
convergence in the presence of noise.
In this paper, we will consider
a normalized gaussian weight function,
\begin{equation}
\label{eq:w_Gaussian}
w(\theta) \equiv \frac{1}{2 \pi \omega^2} e^{-\frac{\theta^2}{2\omega^2}},
\end{equation}
which has convenient analytical properties. In practice,
equation~(\ref{eq:centroid}) can be solved iteratively, by fixing the
weight-function width $\omega$ to an initial estimate of the source
size (see \S\ref{w_choice} for a description of our 
choice of $\omega$ for the HST survey strip).

To characterize the source shape, we consider the weighted multipole
moments of the source intensity,
\begin{equation}
\label{eq:iij_def}
I \equiv \int d^{2}\theta ~w({\mathbf \theta}) i({\mathbf \theta}), ~~
I_{ij} \equiv \int d^{2}\theta ~\theta_{i} \theta_{j} w({\mathbf \theta})
  i({\mathbf \theta}), ~~
I_{ijk} \equiv \int d^{2}\theta ~\theta_{i} \theta_{j} \theta_{k}
  w({\mathbf \theta}) i({\mathbf \theta}), ~~ \mbox{etc.}
\end{equation}
where the origin of the coordinate system was chosen to coincide with
${\mathbf \theta}^{0}$. (By definition, the dipole moment $I_{i}$
vanishes in this coordinate system).  It is usually more convenient to
consider the normalized moments
\begin{equation}
\label{eq:jij_def}
J_{ij} \equiv I_{ij}/I, \  J_{ijk} \equiv I_{ijk}/I,~~\mbox{\rm etc.}.
\end{equation}
The normalized quadrupole moments $J_{ij}$ can be diagonalised as
\begin{equation}
{\bf J} = {\bf R}^{T}(-\alpha) \left(
\begin{array}{cc}
  a^{2} & 0   \\
  0     & b^2 \\
\end{array}
\right) {\bf R}(-\alpha),
\end{equation}
where $a$ and $b$ are the (weighted) major and minor radii, $\alpha$
is the position angle measured counter-clockwise from the positive
$x$-axis, and $^{T}$ stands for the transpose operation.  The rotation
matrix ${\mathbf R}$ is defined as
\begin{equation}
\label{eq:r_matrix}
{\mathbf R}(\varphi) \equiv \left(
\begin{array}{cc}
  \cos(\varphi) & -\sin(\varphi) \\
  \sin(\varphi) & \cos(\varphi) \\
\end{array}
\right).
\end{equation}
Inverting this relation yields
\begin{eqnarray}
(a^{2},b^{2}) & = & \frac{1}{2} \left[ J_{11}+J_{22} \pm
  \sqrt{(J_{11}-J_{22})^{2}+4 J_{12}^{2}} \right] \\
\tan 2 \alpha & = & \frac{2 J_{12}}{J_{11}-J_{22}}.
\end{eqnarray}

\subsection{Rotational Properties}
\label{rotations}
To study the rotational properties of the tensors defined above, let
us consider a new coordinate system which is rotated counter-clockwise
by an angle $\varphi$ from the original positive $x$-axis. In this new
coordinate system, the components $J_{ij}'$ of the normalized
quadrupole moments are related to the unrotated components $J_{ij}$ by
$J_{ij}'=R_{ik}(-\varphi) R_{jl}(-\varphi) J_{kl}$, and similarly for
tensors of higher order, where ${\mathbf R}$ is the rotation matrix
defined in equation~(\ref{eq:r_matrix}). In general, any tensor in 2
dimensions can be decomposed into 2-component (or 1-component for
scalars) $\ell$-spinors $v_{i}$ which rotate as
\begin{equation}
v_{i}'=R_{ij}(- \ell \varphi) v_{j},~~~\ell=0,\pm1,\pm,2,\ldots,
\end{equation}
under this change of coordinates. (These spinors form irreducible
representations of $SO(2)$, the rotation group in 2 dimensions).

For instance, the quadrupole moment $J_{ij}$, which, being symmetric,
consists of 3 independent components, can be decomposed into a scalar
\begin{equation}
\label{eq:d_def}
d^{2} \equiv \frac{1}{2}(J_{11}+J_{22}) = \frac{1}{2} (a^2+b^2),
\end{equation}
which is the mean square radius of the source, and a spin-2 tensor
\begin{equation}
\label{eq:epsilon_def}
\epsilon_{i} \equiv 
\frac{\{ J_{11}-J_{22},2 J_{12} \}}{J_{11}+J_{22}} =
\frac{a^{2}-b^{2}}{a^{2}+b^{2}} \{\cos(2\alpha),\sin(2\alpha)\} 
\end{equation}
which is the ellipticity and has been normalized to follow the
weak-lensing nomenclature. The component $\epsilon_{1}$
($\epsilon_{2}$) corresponds to stretches and compressions parallel to
(at $45^{\circ}$ from) the coordinate axes.

By considering infinitesimal rotations, we can also decompose the
fourth-order moment $J_{ijkl}$, which has 5 independent components. We
find that it can be decomposed into a scalar,
\begin{equation}
\lambda \equiv \left( J_{1111}+2 J_{1122}
  + J_{2222} \right) / (2 d^{2} \omega^{2}),
\end{equation}
a spin-2 tensor,
\begin{eqnarray}
\mu_{1} & \equiv & 
  (-J_{1111} + J_{2222})/
  (2 d^{2} \omega^{2}), \nonumber \\
\mu_{2} & \equiv &
  -2 (J_{1112} + J_{1222})/
  (2 d^{2} \omega^{2}),
\end{eqnarray}
and a spin-4 tensor,
\begin{eqnarray}
\nu_{1} & \equiv &
  (J_{1111}-6 J_{1122}+J_{2222})/
    (2 d^{2} \omega^{2}), \nonumber \\
\nu_{2} & \equiv &
  4 (J_{1112}-J_{1222})/
    (2 d^{2} \omega^{2}).
\end{eqnarray}
For future convenience, these spinors have been normalized with the
weight-function width $\omega$ (Eq.~[\ref{eq:w_Gaussian}]) and the scalar
$d$. These decompositions are useful for simplifying tensors which
are averaged over an ensemble of randomly-oriented galaxies (see
\S\ref{ellip_shear}).

\section{Distortion}
\label{distortion}

In this section, we study the effect of distortions on the source
moments.  In practice, distortions arise from the instrument optics
and from weak lensing. We show how the former can be corrected, and
how the latter can be measured by averaging the ellipticities of an
ensemble of galaxies.

\subsection{Distortion Matrix}
A distortion is an intensity-conserving mapping between the
true position ${\bf x}$ and the observed position ${\bf x'}$ of the
form
\begin{equation}
\label{eq:deltax}
{\bf x'}={\bf x'}({\bf x})={\bf x}+{\bf \delta x}({\bf x}).
\end{equation}
The observed intensity $i'({\bf x})$ is thus related to the
true intensity $i({\bf x})$ by
\begin{equation}
\label{eq:i}
i'({\bf x'})=i({\bf x}({\bf x'})).
\end{equation}
The local properties of the distortion are quantified by the distortion
matrix ${\mathbf \phi}$ which is defined as
\begin{equation}
\label{eq:phi_def}
\phi_{ij} \equiv \left(
\begin{array}{cc}
  \kappa + \gamma_{1} & \gamma_{2}+\rho \\
  \gamma_{2}-\rho & \kappa - \gamma_{1} \\
\end{array}
\right)
\equiv \left. \frac{\partial (\delta x_{i})}{\partial x_{j}}
                 \right|_{{\bf x}^{0}},
\end{equation}
where $\kappa$ is the convergence, $\gamma_{i}$ is the shear, and
$\rho$ the rotation parameter. The convergence $\kappa$ describes
overall rescalings, and the shear components $\gamma_{1}$ and $\gamma_{2}$
describe contractions and dilations parallel to, and at $45^{\circ}$
from, the coordinate axes.  In the case of weak lensing, ${\mathbf
\phi}$ is directly related  to the second-order derivative of the
gravitational potential projected along the line of sight. While the
rotation parameter $\rho$ is expected to be negligible for weak
lensing, it is generally not so for instrumental distortions.

Let us consider a true position ${\bf x}={\bf x}^{0} + {\bf \Delta
x}$, which is close to a reference position ${\bf x}^{0}$ (eg. the
centroid of a source). For small distortions (i.e. if the offset
$\Delta x$ is small compared to the scale on which the distortion
mapping varies), the corresponding distorted position is given by
$x_{i}' = x_{i}^{0 \prime} + \left( \delta_{ij} + \phi_{ij} \right)
\Delta x_{j} + O(\phi^{2})$, where $x_{i}^{0 \prime} \equiv
x_{i}'({\bf x}^{0})$ and $O(\phi^{2})$ denotes higher-order
derivatives of the distortion mapping.  Similarly, the true position
corresponding to a distorted position ${\bf x'}={\bf x}^{0 \prime} +
{\bf \Delta x'}$, is given by $x_{i} = x_{i}^{0} + \left( \delta_{ij}
- \phi_{ij} \right) \Delta x_{j}' + O(\phi^{2})$. By inserting this
expression in equation~(\ref{eq:i}) and by letting $x_{i}^{0
\prime}=x_{i}^{0} \equiv 0$ and $\theta_{i} \equiv \Delta x_{i}'$, we
obtain
\begin{equation}
\label{eq:fprime}
i'(\theta_{i})=i(\theta_{i}-\phi_{ij}\theta_{j}) + O(\phi^{2}).
\end{equation}

\subsection{Effect of Distortion on the Source Moments}
The quadrupole moment $I'_{ij}$ (Eq.~[\ref{eq:iij_def}]) for the
distorted image $i'({\bf \theta})$ (Eq.~[\ref{eq:fprime}]) is given by
\begin{equation}
I_{ij}' = \int d^{2}\theta ~\theta_{i} \theta_{j} w(\theta)
i(\theta_{k} - \phi_{kl} \theta_{l}) + O(\phi^{2}).
\end{equation}
After Taylor expanding and integrating by parts we get, for a Gaussian
weight function (Eq.~[\ref{eq:w_Gaussian}]), $I'_{ij} = I_{ij} + 2
I_{k[i} \phi_{j]k} + I_{ij} \phi_{kk} - \omega^{-2} I_{ijkl} \phi_{kl}
+ O(\phi^{2})$, where the unprimed moments correspond to the
undistorted moments. Similarly, the distorted monopole moment is
related to the undistorted moments by $I' = I + I \phi_{kk} -
\omega^{-2} I_{kl} \phi_{kl} + O(\phi^{2})$.  As a result, the
distorted normalized moments (Eq.~[\ref{eq:jij_def}]) are given by
\begin{equation}
\label{eq:jij_distorted}
J'_{ij} = J_{ij} + D_{ijkl} \phi_{kl} + O(\phi^{2})
\end{equation}
where the distortion susceptibility tensor $D_{ijkl}$ is given by
\begin{equation}
\label{eq:d_tensor}
D_{ijkl} = D_{ijkl}({\bf J}) = 2 \delta_{k[i} J_{j]l} + \omega^{-2} \left(
  J_{ij} J_{kl} - J_{ijkl} \right).
\end{equation}
The brackets denote the symmetrizer, which, for an arbitrary tensor
$A_{i_{1}i_{2} \ldots i_{n}}$ of rank $n$, is defined by
\begin{equation}
\label{eq:symmetrizer}
A_{[i_{1}i_{2} \ldots i_{n}]} \equiv \frac{1}{n!}
\left[A_{i_{1}i_{2} \ldots i_{n}}+ \mbox{\rm all $n!$ permutations of }
\{i_{1},i_{2}, \ldots ,i_{n}\} \right].
\end{equation}
Inverting equation~(\ref{eq:jij_distorted}) yields
\begin{equation}
\label{eq:jij_dcorrection}
J_{ij} = J'_{ij} - D'_{ijkl} \phi_{kl} + O(\phi^{2}),
\end{equation}
where $D'_{ijkl} = D_{ijkl}({\bf J'})$. This expression can be used to
correct the normalized moments for a known distortion.

\subsection{Correction for the Fourth-Order Moments}
\label{jijkl_distortion}
In principle, the correction for the fourth-order moments $J_{ijkl}$
can be derived in a similar way. However, when the weight function
$w(\theta)$ is taken into account, the resulting expressions contain
sixth-order moments and are very cumbersome. As we will see in
\S\ref{HST}, it is sufficient, in practice, to
consider the unweighted ($\omega \rightarrow \infty$) corrections to
$J_{ijkl}$. In this approximation, the corrected fourth-order moments
$J_{ijkl}$ are related to the distorted moments $J_{ijkl}'$ by
\begin{equation}
\label{eq:jijkl_dcorrection}
J_{ijkl} \simeq J_{ijkl}'-4 J_{m[ijk}' \phi_{l]m} + O(\phi^{2}),
\end{equation}
where, as before, the brackets denote the symmetrizer
(Eq.~[\ref{eq:symmetrizer}]).  This expression can be used to correct
$J_{ijkl}'$ for instrumental distortions. 

\subsection{Measurement of the Shear}
\label{ellip_shear}
We now show how the gravitational shear can be measured by averaging
over galaxy ellipticities. For this purpose, we consider galaxy
moments that have been corrected for all instrumental effects,
i.e. from instrumental distortion and PSF convolution, using the
prescriptions presented in the other sections of this paper.  The
effect of a weak general distortion $\phi_{ij}$
(Eq.~[\ref{eq:phi_def}]) on the (corrected) ellipticity of a galaxy
can be derived by substituting the distorted moments $J_{ij}'$
(Eq.~[\ref{eq:jij_distorted}]) into the definition of the ellipticity
$\epsilon_{i}$ (Eq.~[\ref{eq:epsilon_def}]).  This results in a
relation between the distorted ellipticity $\epsilon_{i}$ and the
distortion parameters $\delta_{i} \equiv\{\kappa, \gamma_{1},
\gamma_{2}, \rho \}$ of the form $\epsilon_{i} = A_{ij} \delta_{j} +
O(\phi^{2})$, with
\begin{equation}
A_{ij} \equiv \left(
\begin{array}{cccc}
\mu_{1} + \lambda \epsilon_{1} & 
2 - 2 \epsilon_{1}^{2} - \epsilon_{1} \mu_{1} - 
   \frac{1}{2} \lambda - \frac{1}{2} \nu_{1} &
-2 \epsilon_{1}\epsilon_{2} -\epsilon_{1} \mu_{2} - \frac{1}{2} \nu_{2} &
2 \epsilon_{2} \\
\mu_{2} + \lambda \epsilon_{2} & 
-2 \epsilon_{1}\epsilon_{2} -\epsilon_{2} \mu_{1} - \frac{1}{2} \nu_{2} &
2 - 2 \epsilon_{2}^{2} - \epsilon_{2} \mu_{2} 
   - \frac{1}{2} \lambda + \frac{1}{2} \nu_{1} &
- 2 \epsilon_{1} \\
\end{array} \right),
\end{equation}
where $\lambda, \mu_{i}$ and $\nu_{i}$ are the spinors defined in
\S\ref{rotations}.  Note that $\rho$ is kept here for completeness,
but is expected to vanish for weak lensing distortions.

To measure the shear $\gamma_{i}$, we then average over an ensemble of
galaxies which are assumed to be randomly oriented,
intrinsically. Thus, all we need is the rotational average of the
above relation, which we write as $\langle \epsilon_{i} \rangle=
\langle A_{ij} \rangle \delta_{j} + O(\phi^{2})$. After discarding
all terms which are not rotationally invariant, we obtain
\begin{equation}
\langle A_{ij} \rangle = \left(
\begin{array}{cccc}
0 & G_{1} & -G_{2} & 0 \\
0 & G_{2} & G_{1} & 0 \\
\end{array} \right),
\end{equation}
where
\begin{equation}
G_{1} \equiv 2 - \langle \epsilon^{2} \rangle 
  - \frac{1}{2} \langle \lambda \rangle 
  - \frac{1}{2} \langle {\mathbf \epsilon} \cdot {\mathbf \mu} \rangle,
~~~G_{2} \equiv \frac{1}{2} \langle {\mathbf \epsilon} \times {\mathbf
   \mu} \rangle,
\end{equation}
and $\epsilon^{2} \equiv \epsilon_{1}^{2}+\epsilon_{2}^{2}$,
${\mathbf \epsilon} \cdot {\mathbf \mu} \equiv \epsilon_{1} \mu_{1} +
\epsilon_{2} \mu_{2}$, and ${\mathbf \epsilon} \times {\mathbf \mu}
\equiv \epsilon_{1} \mu_{2} - \epsilon_{2} \mu_{1}$. Note that, to
this order, the convergence $\kappa$ (and the rotation parameter
$\rho$) do not affect the mean ellipticity $\langle \epsilon_{i}
\rangle$.

Since we do not expect the galaxy population to have a preferred
handedness, we can also discard terms which are not parity
invariant. It is easy to check that ${\mathbf \epsilon} \times
{\mathbf \mu}$ is such a term since it changes sign when it is
transformed to a left-handed coordinate system (eg. $x \rightarrow x,
y \rightarrow -y$). We are therefore left with a remarkably simple
relation between the mean ellipticity $\langle \epsilon_{i} \rangle$
and the shear $\gamma_{i}$ given by
\begin{equation}
\label{eq:epsilon_gamma}
\langle \epsilon_{i} \rangle = G \gamma_{i} + O(\phi^{2}),
\end{equation}
where $G \equiv G_{1}$ is the shear susceptibility. This expression
agrees with equation~(B13) in KSB, as corrected by HFKS. The inverse
relation,
\begin{equation}
\label{eq:gamma_epsilon}
\gamma_{i} = E \langle \epsilon_{i} \rangle + O(\phi^{2}),
\end{equation}
where $E\equiv G^{-1}$ can be used to measure the shear from the
averaged ellipticity.

\subsection{Special Cases}
\begin{itemize}
\item Unweighted moments: In this case, the effect of a general
distortion $\phi_{ij}$ (Eq.~[\ref{eq:phi_def}]) on the ellipticity is
more tractable. We find that the distorted ellipticity $\epsilon_{i}'$
is given by
\begin{equation}
\label{eq:epsilon_dist}
\epsilon_{i}' = \epsilon_{i} + 2 (\delta_{ij} - \epsilon_{i} \epsilon_{j})
  \gamma_{j} + 2 e_{ij} \epsilon_{j} \rho + O(\phi^{2}),
\end{equation}
where $\epsilon_{i}$ is the undistorted ellipticity, and the
Levi-Civit\`{a} symbol $e_{ij}$ is defined by $e_{11}=e_{22}=0,
e_{12}=-e_{21}=1$. When averaged over an ensemble of
randomly-distributed sources, this reduces to
\begin{equation}
\langle \epsilon_{i}' \rangle = \left( 2 -
  \langle \epsilon^{2} \rangle \right) \gamma_{i} + O(\phi^2),
\end{equation}
in agreement with equation~(\ref{eq:epsilon_gamma}) in the $\omega
\rightarrow \infty$ limit.

\item Circular source, unweighted moments: If we make the further
simplifying assumption that the undistorted source is circular (${\bf
\epsilon}=0$), the distorted ellipticity (Eq.~[\ref{eq:epsilon_dist}])
becomes
\begin{equation}
\label{eq:epsilon_phi}
\epsilon_{i}' \equiv \epsilon_{i}^{\phi} =  2 \gamma_{i} + O(\phi^{2})
\end{equation}
It is useful to plot the distortion ellipticity ${\bf
\epsilon}^{\phi}$ (rather than $\gamma_{i}$) as a function of position
on the chip, as a measure of the effect of the camera distortion (see
figure \ref{fig:wf_shear}).

\item Radial displacement: Let us consider the case where the
displacement field (Eq.~[\ref{eq:deltax}]) is radial, i.e. where ${\bf
\delta x} = f(x) \frac{\bf x}{x}$, where $x \equiv \left| {\bf x}
\right|$, and $f(x)$ is an arbitrary function. It is easy to show
that, in this case, the distortion tensor (Eq.~[\ref{eq:phi_def}]) is
\begin{equation}
\phi_{ij} = \left(\frac{df}{dx}-\frac{f}{x}\right)
  \frac{x_{i}x_{j}}{x^{2}}
  + \frac{f}{x} \delta_{ij}.
\end{equation}
The corresponding distortion ellipticity (Eq.~[\ref{eq:epsilon_phi}])
is
\begin{equation}
\label{eq:ephi_radial}
{\bf \epsilon}^{\phi} = \left(\frac{df}{dx} - \frac{f}{x} \right)
  {\bf \epsilon}^{x},
\end{equation}
where ${\bf \epsilon}^{x} \equiv \{ x_{1}^{2}-x_{2}^{2}, 2 x_{1} x_{2}
\}/(x_{1}^{2}+x_{2}^{2})$ is the unit radial ellipticity field. From
this expression, it is easy to see that ${\bf \epsilon}^{\phi}$ will be
radial (tangential) if $\left(\frac{df}{dx} - \frac{f}{x} \right)$ is
positive (negative).
\end{itemize}

\section{Convolution}
\label{convolution}
In this section, we study the effect of convolution by a
weakly-anisotropic PSF on the source moments. The PSF can be
decomposed into an isotropic and an anisotropic part. We show how each
part can be corrected for, and how the PSF moments can be derived from
stellar images.

\subsection{Effect of Convolution on the Source Moments}
Let us consider the case where the true galaxy image $i({\bf \theta})$
is convolved by a kernel $p({\bf \theta})$. We take $p({\bf \theta})$
to be normalized so that $\int d^{2}\theta p({\bf \theta}) \equiv 1$,
and centered so that $\int d^{2}\theta ~\theta_{i} p({\bf \theta})
\equiv 0$. The observed image is given by
\begin{equation}
\label{eq:iprime_conv}
i'({\bf \theta}) = \int d^{2}\theta' p({\bf \theta}-{\bf \theta'})
  i({\bf \theta'}),
\end{equation}
so that the observed moments are
\begin{equation}
\label{eq:iij_conv}
I_{ij}' = \int d^{2}\theta \int d^{2}\theta' ~\theta_{i} \theta_{j}
  w({\bf \theta}) p({\bf \theta} - {\bf \theta'}) i({\bf \theta'}).
\end{equation}

There are three angular scales in this equation, namely $\omega$, $g$,
and $a$, corresponding to the size of the weight function $w$, the PSF
kernel $p$, and the (unconvolved) source $i$, respectively.  To
simplify this expression, we need to make an expansion with respect to
the ratio between two of these angular scales. We choose to expand
with respect to $\frac{g}{\omega}$, thereby assuming that the window
function width is much larger than that of the PSF. In practice, the
weight function scale is always chosen to be at least as large as the
source size, i.e.  $\omega \gtrsim a$. To be conservative, we thus
take $a \sim \omega$ to collect the terms in the expansion. KSB
instead, effectively expanded in powers of
$\frac{g}{\sqrt{a^{2}+g^{2}}}$. It is interesting to note that the
expression we derive below for $I_{ij}'$ is nevertheless identical to
theirs, to second order in $\frac{g}{\omega}$.

After a change of variable (${\bf \theta}'' \equiv {\bf \theta} - {\bf
\theta}'$), and a Taylor expansion of $w({\bf\theta}'+{\bf \theta}'')$
about ${\bf \theta}'$, the previous equation becomes $I_{ij}' = I_{ij}
+ I P_{ij} - \frac{2}{\omega^2} I_{k[i}P_{j]k} - \frac{1}{2\omega^{2}}
I_{ij} P_{kk} + \frac{1}{2 \omega^{4}} I_{ijkl}P_{kl} +
O\!\left(\frac{g^4}{\omega^2}\right)$, where the brackets stand for
the symmetrizer (Eq.~[\ref{eq:symmetrizer}]), and $I$ and $I_{ijkl}$
are the (undistorted) weighted moments (Eq.~[\ref{eq:iij_def}]). As
before, we take the weight function $w(\theta)$ to be a normalized
gaussian (Eq.~\ref{eq:w_Gaussian}). The moments $P_{ij}$ are the
(unweighted) moments of the convolution kernel, i.e.
\begin{equation}
\label{eq:pij_def}
P_{ij} \equiv \int d^{2}\theta ~\theta_{i} \theta_{j} p({\bf \theta}).
\end{equation}
As stated above, this agrees, to this order, with KSB, as corrected
for a factor of $\frac{1}{2}$ by HFKS. In \S\ref{stars}, we will
show how the unweighted PSF moments $P_{ij}$ can be derived from
weighted stellar moments. In a similar fashion, it is easy to show
that the convolved monopole moment is $I' = I - \frac{1}{2\omega^2} I
P_{kk} + \frac{1}{2 \omega^4} I_{kl}P_{kl}
+O\!\left(\frac{g^{4}}{\omega^{4}}\right)$.  As a result, the
convolved normalized moments (Eq.~[\ref{eq:jij_def}]) are given by
\begin{equation}
\label{eq:jij_convolved}
J_{ij}' = J_{ij} + C_{ijkl} P_{kl} + O\!\left(\frac{g^{4}}{\omega^{2}}\right),
\end{equation}
where the convolution susceptibility tensor $C_{ijkl}$ is given by
\begin{equation}
\label{eq:c_tensor}
C_{ijkl} = C_{ijkl}({\bf J}) = \delta_{ik} \delta_{jl} - \frac{2}{\omega^2}
  J_{k[i}\delta_{j]l} + \frac{1}{2 \omega^4} \left[ J_{ijkl} - J_{ij} J_{kl}
  \right].
\end{equation}
Inverting these equations yields
\begin{equation}
\label{eq:jij_correction}
J_{ij} = J_{ij}' - C_{ijkl}' P_{kl} +
  O\!\left(\frac{g^4}{\omega^{2}}\right),
\end{equation}
where $C_{ijkl}'=C_{ijkl}({\bf J}')$. The last two equations can be
used to correct the observed moments $J_{ij}'$ for the PSF, and only
require knowledge of the observed quadrupole and fourth-order moments,
$J_{ij}$ and $J_{ijkl}$, of the galaxy, and of the quadrupole moments
$P_{ij}$ of the PSF. We now show how this approximation can be used to
correct for a weakly anisotropic PSF.

\subsection{Correction for the PSF}
\label{psf_decomposition}
In practice, it is often required to include galaxies with size $a$
(and therefore weight-function width $\omega$) only marginally
larger than the PSF size $g$. The PSF correction given by
equation~(\ref{eq:jij_correction}), is thus not directly applicable,
as the expansion series do not converge sufficiently fast.  We can
nevertheless apply the above correction scheme when the PSF 
is weakly anisotropic and sufficiently compact. In this case,
we write the unweighted PSF moments as
\begin{equation}
\label{eq:pij_decomp}
P_{ij} = g^{2} \left(
\begin{array}{cc}
  1 + \epsilon^{p}_{1} & \epsilon^{p}_{2}   \\
  \epsilon^{p}_{2} &  1 - \epsilon^{p}_{1}\\
\end{array}
\right),
\end{equation}
where $g$ is the PSF radius, and $\epsilon_{i}^{p}$ is the PSF
ellipticity, which is assumed to be small. We can then decompose the
kernel $p({\mathbf \theta})$ into the convolution of an isotropic
part $p^{i}({\mathbf \theta})$ with an anisotropic part $p^{a}({\mathbf
\theta})$, as
\begin{equation}
\label{eq:p_decomp}
p=p^{i} \ast p^{a}.
\end{equation}
It is easy to show that this implies that
$P_{ij}=P^{i}_{ij}+P^{a}_{ij}$, where $P^{i}_{ij}$ and $P^{a}_{ij}$
are the unweighted moments of $p^{i}$ and $p^{a}$, respectively.  Without loss
of generality, we further require that
\begin{equation}
\label{eq:pi_Gaussian}
p^{i}({\bf \theta}) \equiv \frac{1}{2\pi
g^2}e^{-\frac{\theta^{2}}{2g^{2}}}
\end{equation}
be a normalized circular Gaussian with standard deviation $g$.  This
implies that 
\begin{equation}
P^{i}_{ij}=g^{2} \delta_{ij},~~\mbox{and}~~ 
P^{a}_{ij}=g^{2} \left(
\begin{array}{cc}
  \epsilon^{p}_{1} & \epsilon^{p}_{2}   \\
  \epsilon^{p}_{2} &  -\epsilon^{p}_{1}\\
\end{array}
\right).
\end{equation}
We now show how to correct each of these components in
turn.

\subsection{Anisotropic Correction}
If we consider a convolution with the anisotropic kernel
$p^{a}({\mathbf \theta})$ alone, the corrected moments
(Eq.~[\ref{eq:jij_correction}]) become
\begin{equation}
\label{eq:jij_acorrection}
J_{ij} = J_{ij}' - C_{ijkl}' P^{a}_{kl} +
  O\!\left(\frac{(g \epsilon^{p})^4}{\omega^{2}}\right),
\end{equation}
where, as before, $C_{ijkl}'=C_{ijkl}({\bf J}')$ is defined in
equation~(\ref{eq:c_tensor}). The residual terms are now supressed by
the factor  $(\epsilon^{p})^4$, and are thus negligible in practice.

\subsection{Isotropic Correction}
As explained earlier, the above approximation can not be applied to
correct for the isotropic part of the PSF.  However, since, by
construction, the isotropic part $p^{i}$ of the kernel is a Gaussian,
we can perform the isotropic correction exactly.  Indeed, by inserting
the form of $p^{i}$ (Eq.~[\ref{eq:pi_Gaussian}]) into
equation~(\ref{eq:iij_conv}) and by integrating twice, we find the
convolved moments to be
$I^{\omega\prime}_{ij} = g_{\omega}^{2} I^{\omega_{g}} \delta_{ij}
+ \left( \frac{g_{\omega}}{g} \right)^{4} I^{\omega_{g}}_{ij}$
where
\begin{equation}
\label{eq:omega_g}
\omega_{g}^{2} \equiv \omega^{2} + g^{2},~~ g_{\omega}^{-2}
\equiv g^{-2} + \omega^{-2},
\end{equation}
and the superscript $^{\omega}$ and $^{\omega_{g}}$ in the moments
$I$, $I_{ij}$, and $I_{ij}'$ indicate the standard deviation of their
respective weight functions. The convolved monopole moment is simply
$I^{\omega\prime}=I^{\omega_{g}}$. Consequently, the convolved
normalized moments are
\begin{equation}
J^{\omega\prime}_{ij} = \left( \frac{g_{\omega}}{g} \right)^{4}
  J^{\omega_{g}}_{ij} + g_{\omega}^{2} \delta_{ij}.
\end{equation}
This convenient relation allows us to relate the convolved moments
$J^{\omega\prime}_{ij}$ with the unconvolved moments
$J^{\omega_{g}}_{ij}$, but this time weighted with the wider weight
standard deviation $\omega_{g}$. For later reference, we explicitly
write the inverse relation
\begin{equation}
\label{eq:jij_icorrection}
J^{\omega_{g}}_{ij} = \left( \frac{g}{g_{\omega}} \right)^{4}
\left( J^{\omega\prime}_{ij} - g_{\omega}^{2} \delta_{ij} \right),
\end{equation}
which can be used to correct for the isotropic part
of the PSF.

\subsection{Correction for the Fourth-Order Moments}
As in the case of distortions (\S\ref{jijkl_distortion}), the general
expression for the correction of the fourth-order moments
$J_{ijkl}$ for convolutions contains sixth-order moments and is
very cumbersome. Here again, it is usually sufficient to consider the
unweighted ($\omega \rightarrow \infty$) corrections to
$J_{ijkl}$. With this approximation, the corrected moments $J_{ijkl}$
are related to the convolved moments $J_{ij}'$ and $J_{ijkl}'$ by
\begin{equation}
\label{eq:jijkl_correction}
J_{ijkl} \simeq J_{ijkl}' - P_{ijkl} - 6 P_{[ij} J_{kl]}' 
  + 6 P_{[ij} P_{kl]},
\end{equation}
where
\begin{equation}
P_{ijkl} \equiv \int d^{2}\theta 
~\theta_{i}\theta_{j}\theta_{k}\theta_{l} p({\bf \theta})
\label{eq:pijkl_def}
\end{equation}
is the (unweighted) fourth-order PSF moment,
and the brackets denote the symmetrizer
(Eq.~[\ref{eq:symmetrizer}]). This expression can be used to correct
$J_{ijkl}'$ for convolutions. In the next section, we will show how
$P_{ijkl}$ can be estimated from stellar moments.

\subsection{Measurement of the PSF with Stars}
\label{stars}
Stars are point-like and therefore have an intensity profile given by
$i^{*}({\mathbf \theta}) = S \delta^{(2)}({\mathbf \theta})$, where
$S$ is the flux. The intrinsic moments (Eq.~[\ref{eq:iij_def}]) of a
star are thus $I^{*}=S w_{*}(0)$ and $I^{*}_{ij}=J^{*}_{ij}=0$.  We
allow for the possiblity that the weight function, $w_{*}({\bf
\theta})$, for the stars be different from that for galaxies, $w({\bf
\theta})$. After convolution (see Eq.~[\ref{eq:jij_correction}] and
above), the moments become $I^{* \prime} = I^{*} - \frac{1}{2
\omega_{*}^{2}} I^{*} P_{kk} +
O\!\left(\frac{g^{4}}{\omega_{*}^{4}}\right)$ and $I^{*
\prime}_{ij}=I^{*} P_{ij} - \frac{1}{2 \omega_{*}^{2}} I^{*} P_{ijkk}
+ O\!\left( \frac{g^{6}}{\omega_{*}^{4}} \right)$, where $P_{ijkl}$
was defined in equation~(\ref{eq:pijkl_def}).  As a result, the
normalized moments become
\begin{equation}
\label{eq:jij_star}
J_{ij}^{* \prime} = P_{ij} + \frac{1}{2 \omega_{*}^{2}}
  \left[ P_{ij} P_{kk} - P_{ijkk} \right] +
  O\!\left( \frac{g^{6}}{\omega_{*}^{4}} \right) .
\end{equation}
It is also easy to show that the observed fourth-order moments are
given by $I_{ijkl}^{*\prime} = I^{*} P_{ijkl} + O\!\left
( \frac{g^{6}}{\omega_{*}^{2}} \right)$, while their normalized
version is
\begin{equation}
J_{ijkl}^{*\prime} = P_{ijkl} + 
  O\!\left( \frac{g^{6}}{\omega_{*}^{2}} \right).
\end{equation}
We can invert these equations to obtain
\begin{equation}
\label{eq:pij_correction}
P_{ij} = J_{ij}^{*\prime} - \frac{1}{2 \omega_{*}^{2}}
  \left[ J_{ij}^{*\prime} J_{kk}^{*\prime} - J_{ijkk}^{*\prime} \right] +
  O\!\left( \frac{g^{6}}{\omega_{*}^{4}} \right),
\end{equation}
and
\begin{equation}
\label{eq:pijkl_correction}
P_{ijkl} = J_{ijkl}^{*} + O\!\left( \frac{g^{6}}{\omega_{*}^{2}}
\right).
\end{equation}

With these expressions, the unweighted PSF moments $P_{ij}$ and
$P_{ijkl}$ can be derived from the observed stellar moments
$J_{ij}^{*\prime}$ and $J_{ijkk}^{*\prime}$. These can then be
corrected for the camera distortion using
equations~(\ref{eq:jij_dcorrection},\ref{eq:jijkl_dcorrection}) with
$\omega \rightarrow \infty$, and then used in
equations~(\ref{eq:jij_acorrection},\ref{eq:jij_icorrection},\ref{eq:jijkl_correction})
to correct the galaxy moments. For this purpose, $P_{ij}$ and
$P_{ijkl}$ need to be interpolated across the chip. In practice, this
can be done by fitting a low order polynomial to each component
separately. Figure~\ref{fig:m4sm} shows the PSF ellipticities for the
WFPC2 camera derived from globular cluster observations
(Figures~\ref{fig:m4ellipt} and \ref{fig:ngc6572}), after correction and
interpolation.

If the star and galaxy weight functions are equal
($\omega=\omega_{*}$), then one only needs to keep the first term in
equation~(\ref{eq:pij_correction}).  However, more accurate
measurements of the stellar shapes can be achieved by taking a
narrower weight function. In addition, it is desirable to avoid
recomputing the stellar moments for each value of $\omega$, which is
often taken to be a function of the galaxy size.  One would then
choose $\omega_{*} < \omega$, making
equation~(\ref{eq:pij_correction}) converge more slowly than
equation~(\ref{eq:jij_correction}). In this case, one would thus need
to keep the second term in equation~(\ref{eq:pij_correction}) and to
ensure that the residual error ($O\!\left(
\frac{\rho^{6}}{\omega_{*}^{4}} \right)$) is acceptable compared to
that for equation~(\ref{eq:jij_correction}) ($O\!\left(
\frac{\rho^{4}}{\omega^{2}} \right)$).

\subsection{Special Cases}

\begin{itemize}

\item Unweighted moments: In this case, the convolved moments become
(Eq.~[\ref{eq:jij_convolved}] with $\omega \rightarrow \infty$)
$J_{ij}^{\prime} = J_{ij} + P_{ij}$. Thus, the observed square radius is
\begin{equation}
d^{\prime 2} = d^{2}+g^{2},
\label{eq:d_obs}
\end{equation}
and the observed ellipticity is $\epsilon_{i}'= \frac{d^{2}
\epsilon_{i} + g^{2} \epsilon_{i}^{p}}{d^{2}+g^{2}}$, where $g^{2}$
and $\epsilon_{i}^{p}$ are defined by equation~(\ref{eq:pij_decomp}).
It is sometimes useful to consider the moments $J_{ij}^{\rm deconv}
\equiv J_{ij} - g^{2} \delta_{ij}$, which have been deconvolved from
the isotropic part of the PSF only. The associated radius is $d^{\rm
deconv}=d$, while the associated ellipticity is $\epsilon_{i}^{\rm
deconv} = \epsilon_{i} + \left( \frac{g}{d} \right)^{2} \epsilon_{i}^{p}$.

\item Unweighted moments, circular source: In this case,
$\epsilon_{i}=0$ and thus the observed ellipticity becomes
\begin{equation}
\label{eq:fd}
\epsilon_{i}' \equiv f_{d}~ \epsilon_{i}^{p}
  =\frac{g^2}{d^{2}+g^{2}}\epsilon_{i}^{p},
\end{equation}
while the isotropically deconvolved ellipticity becomes
\begin{equation}
\label{eq:fd_deconv}
\epsilon_{i}^{\rm deconv} \equiv f_{d}^{\rm deconv}~\epsilon_{i}^{p} =
  \left( \frac{g}{d} \right)^{2} \epsilon_{i}^{p}.
\end{equation}
We have introduced the pre- and post-deconvolution reduction factors
$f_{d}$ and $f_d^{\rm deconv}$. These expressions are useful to
estimate the effect of the PSF anisotropy on a galaxy.

\end{itemize}

\section{Combined Effect of Distortion and Convolution}
\label{comb}
In practice, an image is deformed by a series of distortions and
convolutions. The full treatment of the combined effect of a
distortion and convolution is impractical given the complexity of the
$D_{ijkl}$ and $C_{ijkl}$ tensors
(Eqs.~[\ref{eq:c_tensor},\ref{eq:d_tensor}] ). However, the general
behavior of the ellipticity is captured by considering the simplified
case of a circular unweighted source which is  subject to a
weak distortion.  This provides a good model of ellipticities observed
in practice (see \S\ref{Strip} below).

In the unweighted case ($w(\theta)=1$) , the normalized quadrupole
moments $J_{ij}'$ of an image deformed by a convolution followed by a
distortion are (see
Eqs.~[\ref{eq:jij_correction},\ref{eq:jij_dcorrection}] with $\omega
\rightarrow \infty$)
\begin{equation}
J_{ij}'= (J_{ij}+P_{ij}) + (J_{ik}+P_{ik}) \phi_{kj}
  + (J_{jk}+P_{jk}) \phi_{ki} + O(\phi^{2}),
\end{equation}
where $P_{ij}$ are the PSF moments (Eq.~[\ref{eq:pij_def}]),
$\phi_{ij}$ is the distortion matrix (Eq.~[\ref{eq:phi_def}]), and
$J_{ij}$ are the undeformed moments.  We will assume that the PSF is
weakly anisotropic, i.e. that $\epsilon_{i}^{p}$ defined in
equation~(\ref{eq:pij_decomp}) is small. Then, for an intrinsically
circular source ($J_{ij}=d^{2} \delta_{ij}$), the ellipticity
$\epsilon_{i}'$ of the deformed source is
\begin{equation}
\epsilon_{i}'=\frac{d^{2}2\gamma_{i}+g^{2}(\epsilon_{i}^{p}+2\gamma_{i})}
  {d^{2}+g^{2}} + O((\phi,\epsilon^{p})^2),
\end{equation}
where the deformation matrix $\phi_{ij}$ was parametrized as in
equation~(\ref{eq:phi_def}), and $g$ is the PSF radius defined in
equation~(\ref{eq:pij_decomp}).  Notice that the convergence $\kappa$
and rotation parameter $\rho$ do not appear in this first order
expression. The observed square radius $d^{\prime 2} \equiv
(J_{11}'+J_{22}')/2$ is given by
\begin{equation}
d^{\prime 2} = (1+2 \kappa) (d^{2}+g^{2}) + O((\phi,\epsilon^{p})^2).
\end{equation}
For stars ($d=0$), the observed ellipticity $\epsilon_{i}^{*}$ becomes
\begin{equation}
\epsilon_{i}^{*} = \epsilon_{i}^{p} + 2 \gamma_{i} + O((\phi,\epsilon^{p})^2).
\end{equation}
We can thus rewrite $\epsilon_{i}'$ in terms of observables as
\begin{equation}
\epsilon_{i}' = \left[ 1-\left( \frac{g}{d'} \right)^{2} \right] 2 \gamma_{i}
  + \left( \frac{g}{d'} \right)^{2} \epsilon_{i}^{*}
  + O((\phi,\epsilon^{p})^2) .
\label{eq:comb}
\end{equation}
This expression is useful for computing the expected deformation of
objects as a function of their observed size. This simplified model
agrees well with the observed deformations in the Survey Strip
(see \S\ref{Strip}).

\section{Validity of the Method}
\label{validity}

\subsection{Shortcomings}
\label{shortcomings}
As was recently pointed out (\markcite{kai99}Kaiser 1999;
\markcite{kui99} Kuijken 1999), the KSB method has several
shortcomings\footnote{We thank Nick Kaiser, the referee, for pointing
out and clarifying these shortcomings.} which limit its validity and
accuracy. First, the method requires a decomposition of the PSF into
the convolution of an isotropic part and a compact anisotropic
part. For most PSFs encountered in practice, this decomposition is
formally ill-defined and the anisotropic part is not necessarily
sufficiently compact. Another problem results from the fact that most
PSFs do not fall off fast enough for the second moments (and higher
moments) to converge.  Consequently, the weighted second moments of
the PSF depend strongly on the size of the window function. These
problems are particularly severe for the PSF of HST which has broad
wings extending beyond a central core, and thus cast doubts on the
validity of the numerous weak-lensing analyses based on HST
observations.

Since our method is based on the same principles as that of the KSB
method, it also suffers from the same shortcomings. In addition, we
have made the further assumption that, in the decomposition of the
PSF, the isotropic part can be taken to be a gaussian (see
\S\ref{psf_decomposition}). This has the advantage of greatly
simplifying the deconvolution, but can arguably produce further
inaccuracies.

These problems are certainly worrisome, and might eventually be solved
by considering different methods such as those proposed by
\markcite{kai99}Kaiser (1999) and \markcite{kui99}Kuijken (1999).
However, the KSB approach has the advantage of being relatively simple
and of requiring only a small amount of information about the PSF and
galaxy shapes, namely the first few multipole moments. Moreover, these
shortcomings are mainly formal in nature.  In practice, one indeed
always measures moments with a weight function width that is close to
the width of the central core of the PSF, and which therefore do not
diverge. For instance, for the HST PSF, the secondary sidelobes which
compose the extended wings are smaller in amplitude 
than the central core by about
one order of magnitude. Since the central core dominates
in the convolution, one therefore expects that the corrections will be
approximately correct.  In the following paragraphs, we describe
numerical simulations designed to test this assertion quantitatively.

\subsection{Simulations}
To test the weak link in our method, namely the PSF correction, we
performed a series of numerical simulations.  We first generated a
WFPC2 PSF using the Tiny Tim software (\markcite{kri97}Krist \& Hook
1997). We chose the PSF to be particularly non-circular by placing it
at the lower-right-hand corner of chip 2, at pixel coordinate
(100,100) (see \S\ref{HST} for a description of the WFPC2 camera and
of its PSF).  We oversampled the PSF by using pixels which are 10 times
smaller than the WFPC2 pixels, and did not add any noise to the
images. This isolates the systematic deconvolution errors from the
random errors produced by pixelization and noise. To maintain
uniformity, we will nevertheless quote angular sizes in WFPC2 pixels.

We then measured the weighted moments $P_{ij}$ of the PSF as described
in \S\ref{stars}, using a range of weight function widths
$\omega_{*}$.  For $\omega_{*}=2$ pixels, we found the PSF size and
ellipticity components (Eq.~[\ref{eq:pij_decomp}]) to be $g \simeq
0.86$ pixels, and $\epsilon_{i} \simeq \{-.009,-.048\}$. As expected,
we found both $g$ and $\epsilon_{i}$ to diverge as $\omega_{*}$
increases. Fo the rest of the simulations, we set $\omega_{*}=2$
pixels, which is the stellar weight width that we used in our analysis
of the survey strip (see \S\ref{w_choice}).

We then convolved the PSF with elliptical gaussian ``galaxies'' of
various intrinsic sizes $d$ and ellipticities $\epsilon_{i}$. Since
the source of the problem lies with the PSF shape, this simplified
galaxy model is sufficient for our purposes. We measured the moments
of the convolved gaussian using a weight function width of
$\omega=\mbox{max}(2,d)$ pixels, just as we did for the analysis of
the survey strip (\S\ref{w_choice}). We then applied our PSF
correction method (\S\ref{convolution}) to obtain the corrected
ellipticity $\epsilon_{i}^{\mbox{\rm corrected}}$, weighted with an
effective width $\omega_{g}$ (Eq.~[\ref{eq:omega_g}]).  For an
elliptical gaussian source, it is easy to show that the true weighted
ellipticity is $\epsilon_{i}^{\mbox{\rm true}}= \epsilon_{i} \left[ 1
+ \left( \frac{d}{\omega_{g}}\right)^{2} (1- \epsilon^{2})
\right]^{-1}$. The error $\Delta \epsilon_{i}$ induced by the PSF
correction on the galaxy ellipticity is thus
\begin{equation}
\Delta \epsilon_{i} \equiv \epsilon_{i}^{\mbox{\rm corrected}}
  - \epsilon_{i}^{\mbox{\rm true}}.
\end{equation}
We computed this residual error for a range of intrinsic galaxy
sizes and ellipticities. The results of these simulations are 
presented in the next section. 

\subsection{Results}
\begin{figure}
\plotone{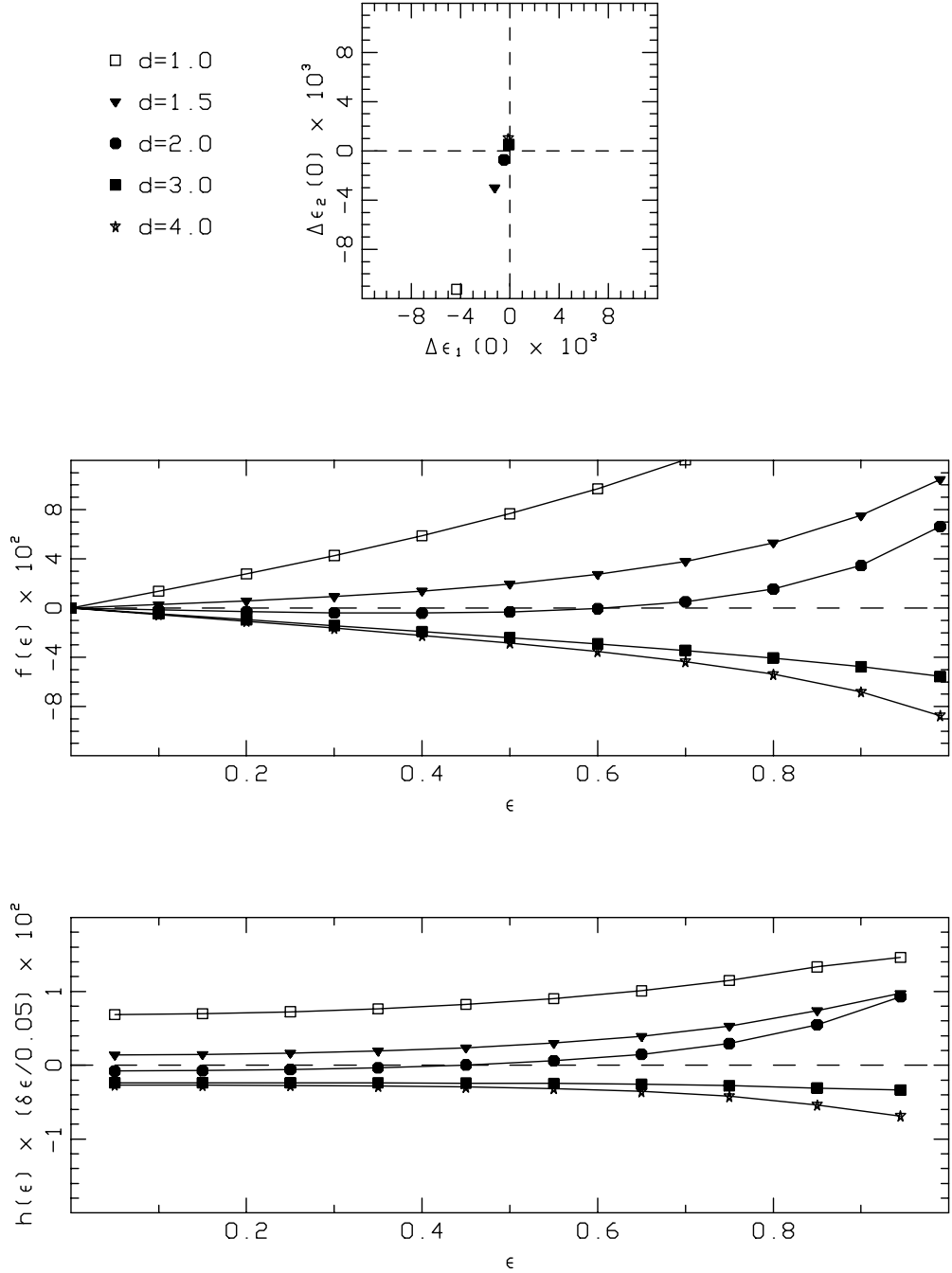}
\caption{Residual errors of the PSF correction determined from
numerical simulations. Panels (a), (b) show, respectively, the
constant term $\Delta \epsilon_{i} (0)$ and the radial function
$f(\epsilon)$, for measuring absolute ellipticities. Panel (c) shows
the radial function $h(\epsilon)$ for measuring an ellipticity change
of $\delta \epsilon_{i}=0.05$ (see text). These quantities are plotted
as a function of the unweighted galaxy ellipticity $\epsilon$ and size
$d$ (in WPC2 pixels).}
\label{fig:errors}
\end{figure}

\begin{figure}
\epsscale{1.0}
\plotone{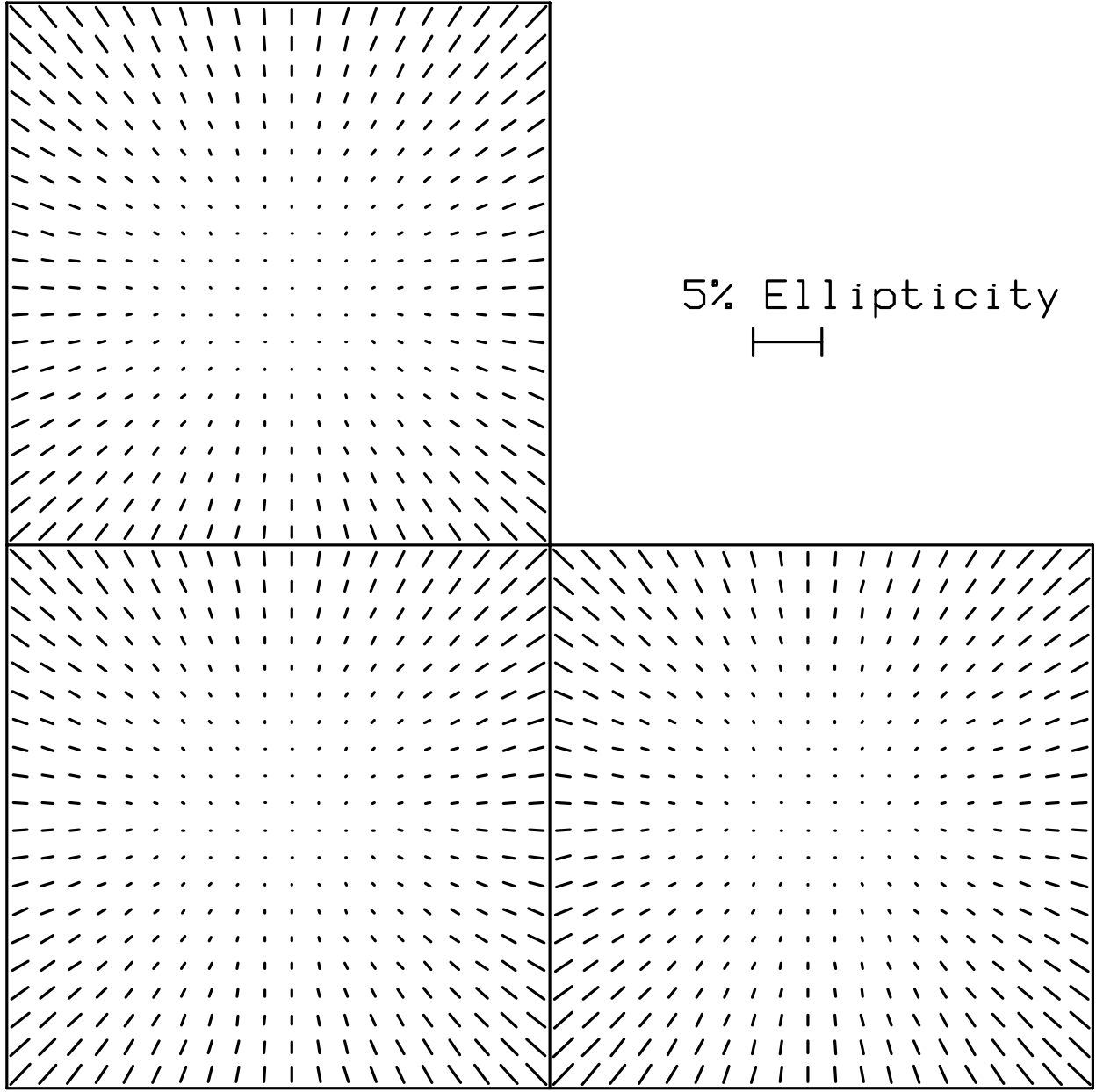}
\caption{The WFPC2 Detector Shear pattern. The quantity plotted is
$\epsilon^{\phi}_{i}= 2 \gamma_{i}$ which is the ellipticity induced
by the detector shear for an intrinsically circular source.  Chips 2,
3, and 4 are in the upper-left, lower-left, and lower-right
corners, respectively. Chip 1 (the Planetary Camera) is not shown. Each
chip is 1.3 arcmin on a side.}
\label{fig:wf_shear}
\end{figure}

For the range of galaxy shapes which we considered ($1.0<d<4.0$ pixels
and $\epsilon_{i}<1$), we found that, for a given galaxy size $d$, the
ellipticity error $\Delta \epsilon_{i}$ is very well approximated by
\begin{equation}
\label{eq:error_model}
\Delta \epsilon_{i} \simeq \Delta \epsilon_{i}(0) + f(\epsilon)
\hat{\epsilon_{i}},
\end{equation}
where $\Delta \epsilon_{i}(0)$ is a constant, $\hat{\epsilon_{i}}
\equiv \epsilon_{i}/\epsilon$ is the unit radial ellipticity vector,
and $f(\epsilon)$ is a function describing the radial behavior.  The
constant term $\Delta \epsilon_{i}(0)$ is shown in
figure~\ref{fig:errors}a for several relevant galaxy sizes, and has a
modulus less than .004 for $d>1.5$ pixels. Figure~\ref{fig:errors}b
shows the radial term $f(\epsilon)$ as a function of the unweighted
galaxy ellipticity modulus $\epsilon$, for the same galaxy sizes. The
radial term depends strongly on the galaxy size $d$ and ellipticity
$\epsilon$, and has an amplitude of several percent for moderate
ellipticities ($\epsilon<0.6$) and sizes ($d>1.5$ pixels). This
amplitude is the size of the systematic error made in measuring the
ellipticity of a single galaxy.

However, for weak lensing measurements, one is not eventually
interested in the ellipticity of a single galaxy, but in measuring a
global change in ellipticity averaged over a galaxy ensemble. Let us
consider a change in ellipticity $\epsilon_{i} \rightarrow
\epsilon_{i}^{\prime}= \epsilon_{i}+\delta_{i}$ produced by
lensing\footnote{Strictly speaking, the ellipticity change
$\delta_{i}$ produced by weak lensing has a weak dependence on
$\epsilon_{i}$ (See Eq.[\ref{eq:epsilon_dist}]). While this effect is
only important for large ellipticities, the ensemble average is
dominated by small ellipticities. For the purpose of estimating the
size of the average correction error, we can thus ignore this small
effect and consider a constant ellipticity shift.}.  The error in the
measurement of $\delta \epsilon_{i}$ is
\begin{equation}
\Delta \delta \epsilon_{i}
\equiv \left\langle \Delta \epsilon_{i} (\epsilon_{i}') \right\rangle
\simeq \left\langle \Delta \epsilon_{i} \right\rangle +
       \left\langle \frac{\partial \Delta \epsilon_{i}}{\partial
        \epsilon_{i}} \right\rangle \delta \epsilon_{i},
\end{equation}
where the brackets refer to the average over the galaxy ensemble.
For the functional form of equation~(\ref{eq:error_model}), this
becomes
\begin{equation}
\label{eq:delta_delta_e}
\Delta \delta \epsilon_{i} \simeq \Delta \epsilon_{i}(0)
  + h(\epsilon) \delta \epsilon_{i},
\end{equation}
where $h(\epsilon) \equiv \frac{1}{2} \left[ f'(\epsilon) +
\frac{f(\epsilon)}{\epsilon} \right] $. The first term was plotted on
figure~\ref{fig:errors}a. The second term is shown as a function of
$\epsilon$ and $d$ in figure~\ref{fig:errors}c, for an ellipticity
change of $\delta \epsilon_{i} = 0.05$. Since, for our sample $G \simeq
1.4$ (Eq.~[\ref{eq:epsilon_gamma}]), this value of $\delta \epsilon_{i}$
corresponds to a shear of $\gamma_{i} \simeq .04$ and is therefore
representative of shear signals expected from large-scale
structure. For moderate ellipticities ($\epsilon<0.6$) and sizes
($d>1.5$ pixels), the second term has an amplitude less than 0.004.

Interestingly, both terms in equation~(\ref{eq:delta_delta_e}) tend to
average out, when the distribution of galaxy sizes is considered.
Indeed, in our treatment of the survey strip (see \S\ref{budget}), we
selected a sample of ``large'' galaxies with $d>1.5$ pixels (and
$\epsilon<1$, which is not always satisfied because of noise). The
average size for this sample is $\langle d \rangle \simeq 2.58$
pixels, which is close to the value of $d$ for which both terms change
sign (see figure~\ref{fig:errors}). The ellipticity dispersion for
this sample is $\sigma_{\epsilon} \simeq .31$. Taking $\langle d
\rangle$ and $\sigma_{\epsilon}$ to be representative for this sample,
we find the average residual errors to be $\Delta \delta \epsilon_{i}
\approx 0.001$. This cancellation may, however, depend on the galaxy
profile, while we have only considered gaussians in these
simulations. To be conservative, we therefore take the residual error
for our sample to be about 0.004, which is the error before averaging
over $d$ values. This is close to the residual error estimated from
the anisotropic correction of the stars in our analysis of globular
cluster fields (\S\ref{budget}).

We also studied how the choice of the weight function widths $\omega$
and $\omega_{*}$ affect these results. We found that for moderate
values ($1<\omega_{*}<3$ and $0.5 d< \omega < 2 d$), the magnitude of
the residual errors did not change significantly from the
above. However, the errors were much larger for $\omega$ and
$\omega_{*}$ outside of these ranges. This is, of course, a
consequence of the divergences present in the PSF moments.

We conclude that, the ellipticity error produced by the PSF correction
can be several percent for an individual galaxy observed with
WFPC2. However, it is only about 0.004 when averaged over a galaxy
ensemble with $d>1.5$ pixels, provided moderate values of the weight
function widths are used. This does not alleviate the serious formal
shortcomings of the method, and the need for a search for more robust
methods (see \markcite{kai99}Kaiser (1999); \markcite{kui99}Kuijken
(1999)).

\section{Application to HST images}
\label{HST}
As an application of our method, we consider weak-lensing measurements
with the WFPC2 camera onboard HST. The camera consists of 3 $800
\times 800$ pixel chips, with a pixel size of 0.1 arcsec.  With the
F814W filter, the PSF has a FWHM of about 0.09 arcsec. The
instrumental effects for WFPC2 have been studied in detail by
HFKS. They showed that the two main systematic effects for this
instrument are the camera distortion and the PSF. We repeat their
analysis by considering each of these effects in turn. We then apply
these results to the galaxies found in the Survey Strip.

\subsection{Camera Distortion}
\label{hst_distortion}
The camera distortion is caused mainly by the flat fielder, a refractive
element in the WFPC2. This effect has been studied by Holtzman et
al. (1995),\markcite{hol95} who quantified this effect using
astrometric measurements of globular cluster observations. They
modeled the distortion by a cubic polynomial in chip position for
each chip. The polynomial coefficients can be used to derive a
transformation matrix $\phi^{\rm camera}_{ij}$ between the WFPC2 chip
coordinates (which are the distorted coordinates ${\bf x'}$ in
Eq.~[\ref{eq:deltax}]) and the undistorted coordinates (${\bf x}$
in the same equation). 

The distortion field is shown in figure~\ref{fig:wf_shear}.  The
quantity plotted is $\epsilon^{\phi}_{i} \equiv 2 \gamma_{i}$, the
ellipticity induced by the distortion field $\phi_{ij}$ on an
intrinsically circular source (see Eq.~[\ref{eq:epsilon_phi}]). The
pattern is radial and increases in magnitude with distance from the
chip center. This differs from the results of HFKS who derived a
tangential rather than radial camera distortion field. The radial
nature of the pattern is confirmed by the fact that, using the
coefficients of Holtzman et al. (1995) in
equation~(\ref{eq:ephi_radial}), we find $\left(\frac{df}{dx} -
\frac{f}{x} \right) > 0$. (This can also be confirmed by drawing a box
connecting the heads and tails of four adjacent arrows in Figure 15 in
Holtzman et al.)
\label{hfks_camera}

To study the profile of the ellipticity pattern, it is useful to
define the rotated ellipticity by
\begin{equation}
\epsilon_{i}^{r} \equiv R_{ij}(-2\varphi)~\epsilon_{j},
\end{equation}
where $\epsilon_{i}$ is the ellipticity of an object in the chip
frame, $R_{ij}$ is the rotation matrix defined in
equation~\ref{eq:r_matrix}, and $\varphi$ is the polar angle of the
source position about the center of the chip, measured
counter-clockwise from the x-axis. A positive (negative) value of
$\epsilon_{1}^{r}$ corresponds to a radial (tangential) ellipticity
pattern. A positive (negative) value of $\epsilon_{2}^{r}$ corresponds
to an anti-clockwise (clockwise) swirl pattern. The meaning of
$\epsilon_{i}^{r}$ is illustrated in figure~\ref{fig:e_rotate}.

The profile of the rotated distortion ellipticity is shown as the
dot-dashed line on figure~\ref{fig:inst}. The mean ellipticity
averaged over the 3 chips is listed in table~\ref{tab:stats}. The mean
rotated ellipticities are found to be $\langle \epsilon_{1}^{r} \rangle
\simeq 0.007$ and $\langle \epsilon_{2}^{r} \rangle < 0.001$, while
the mean absolute ellipticity $\langle \epsilon_{i} \rangle$ is less
than 0.001. The effect of the camera distortion is thus small, but
nevertheless comparable to the lensing signal expected from
large-scale structure.

\subsection{Point-Spread Function}
\label{hst_psf}
The PSF is affected by diffraction by the telescope and scattering of
light from different parts of the telescope and the optics.  The PSF
depends both on wavelength and time.  The time dependence has two
components (Krist and Hook, 1997)\markcite{kri97}.  The first one is
due to orbit-to-orbit ``breathing'' of the telescope.  As the
telescope orbits the earth every 90 minutes, it passes into and out of
sunlight. The heating and cooling of the telescope changes its size,
and thus its focus and PSF. The second time-dependent factor arises
from a change in the focus of the telescope over longer periods of
time. This change is produced by the outgassing of the graphite epoxy
truss which supports the primary mirror.  Approximately every six
months, the secondary mirror is moved to bring the telescope back into
optimal focus. The time-dependence of the focus position is summarized
in Figure~\ref{fig:focus}. 

\subsubsection{Tiny Tim}
\label{Tiny Tim}
As a first step, we used the Tiny Tim software to model the PSF of the
HST (Krist \& Hook 1997\markcite{kri97}). Tiny Tim takes into account
the diffraction of light by the telescope to create a PSF, given the
instrument (WFPC2 in our case), chip number, chip position,
filter, spectrum of the object being modeled, and the focus of the
telescope.  Tiny Tim does not include the geometric distortion
discussed above.

We used Tiny Tim to model the PSF across the three WFPC2 chips.  We
chose the F814W filter and the spectrum of the objects to be similar
to that of a type G star (B$-$V=0.619), which is typical of galaxies in
the Survey Strip (see \S\ref{Strip}). We measured the moments
(Eq.~[\ref{eq:jij_def}]) and weighted them with a circular Gaussian
(Eq.~[\ref{eq:w_Gaussian}]). For all stellar measurements, we adopted
a weight-function width of $\omega_{*}=2$ pixels, which resulted in an
optimal sensitivity.

\begin{figure}
\plotone{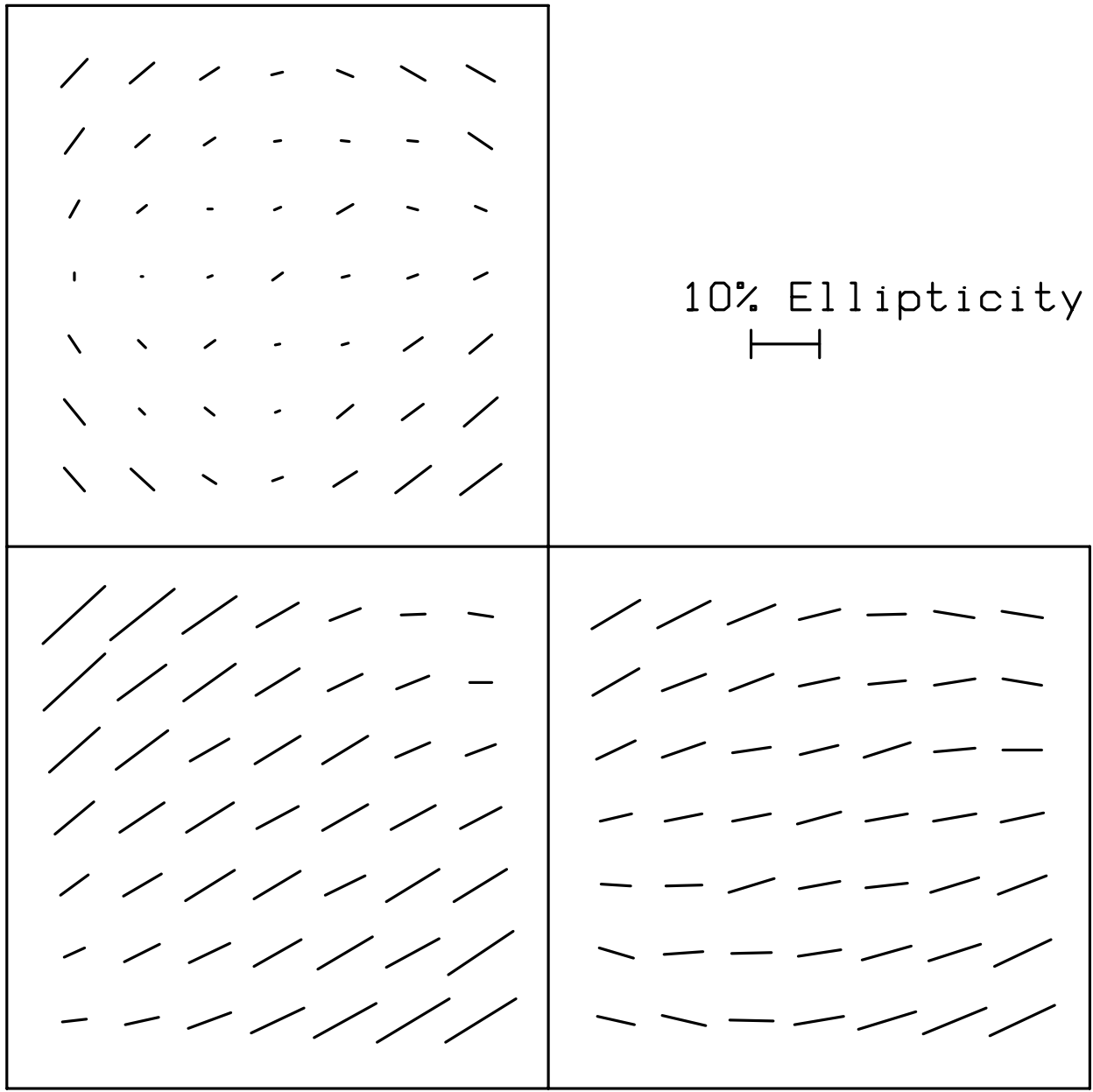}
\caption{PSF ellipticity predicted by Tiny Tim. This figure 
corresponds to a focus of $-1.5\mu$m, close to the mean
value for the Survey Strip observations. The ellipticities
are shown after application of the camera distortion.}
\label{fig:tt}
\end{figure}

\begin{figure}
\plotone{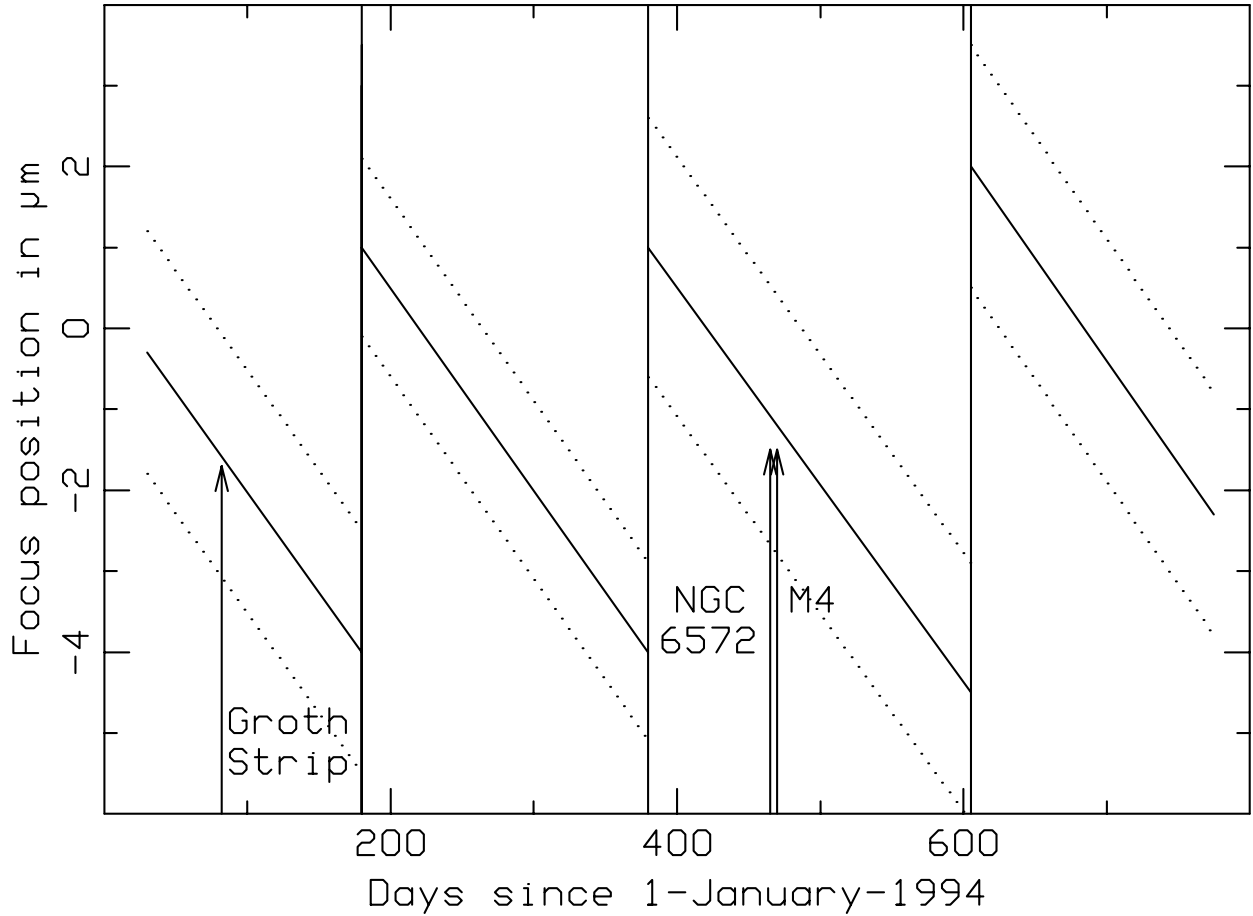}
\caption{HST focus as a function of time. The vertical lines give the
times of mirror movements generated to bring the telescope back into
focus. The diagonal lines represent the average mirror position as a
function of time. Orbit-to-orbit ``breathing'' produces considerable
variations (up to several microns) about these mean values.  The {\it
rms} value of these variations is shown by the dotted lines.  This
figure is patterned after Figure $5.7$ in Biretta (1996). The values
corresponding to the observations of the Survey Strip, and of the M4
and NGC6572 globular clusters are also indicated.}
\label{fig:focus}
\end{figure}

\begin{figure}
\plotone{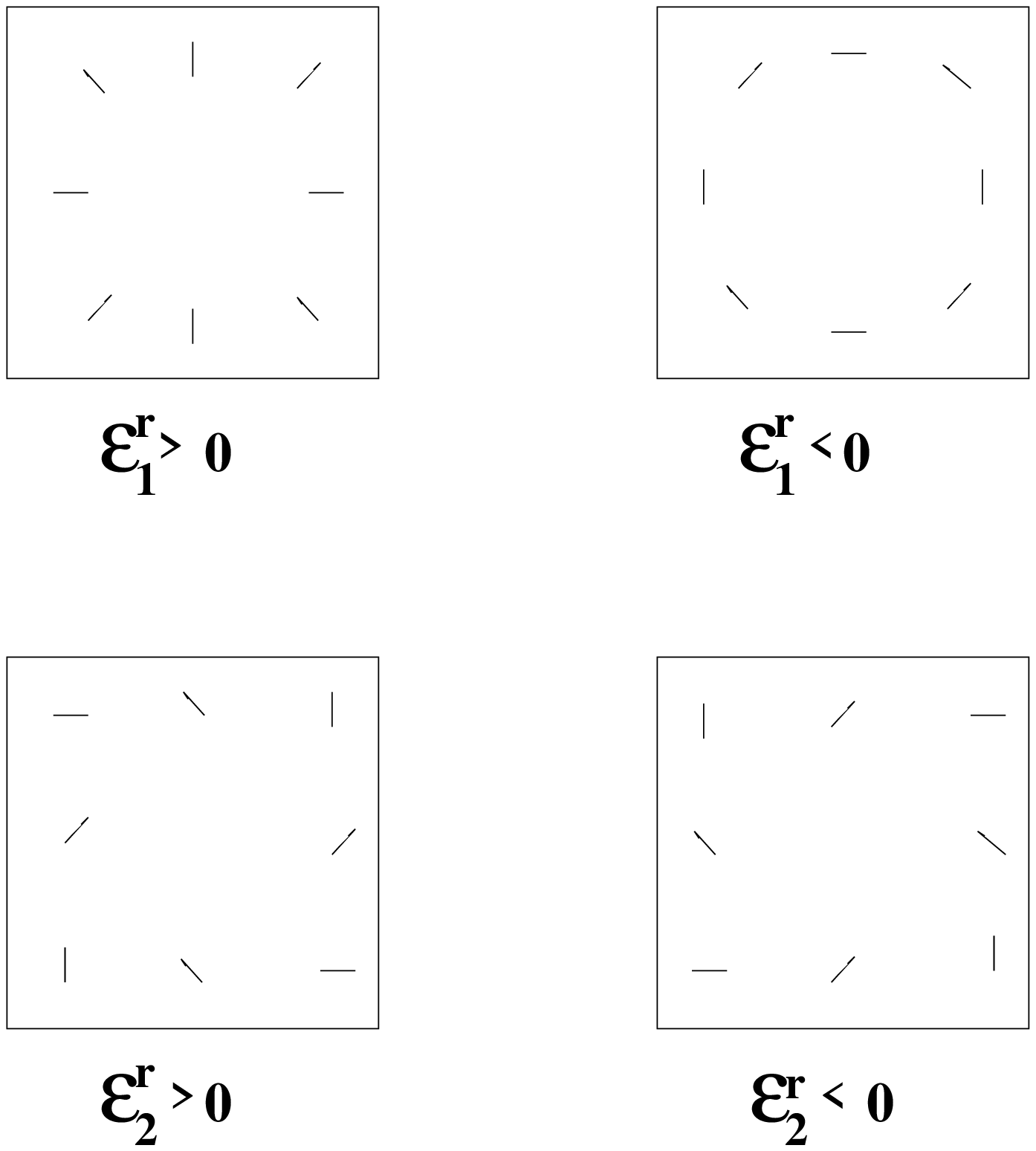}
\caption{Illustration of the meaning of rotated ellipticity
$\epsilon_{i}^{r}$. This ellipticity is defined by choosing
coordinates axes which are rotated about the chip center.
The patterns corresponding to positive and negative values for
each component of $\epsilon_{i}^{r}$ are shown.}
\label{fig:e_rotate}
\end{figure}

\begin{figure}
\plotone{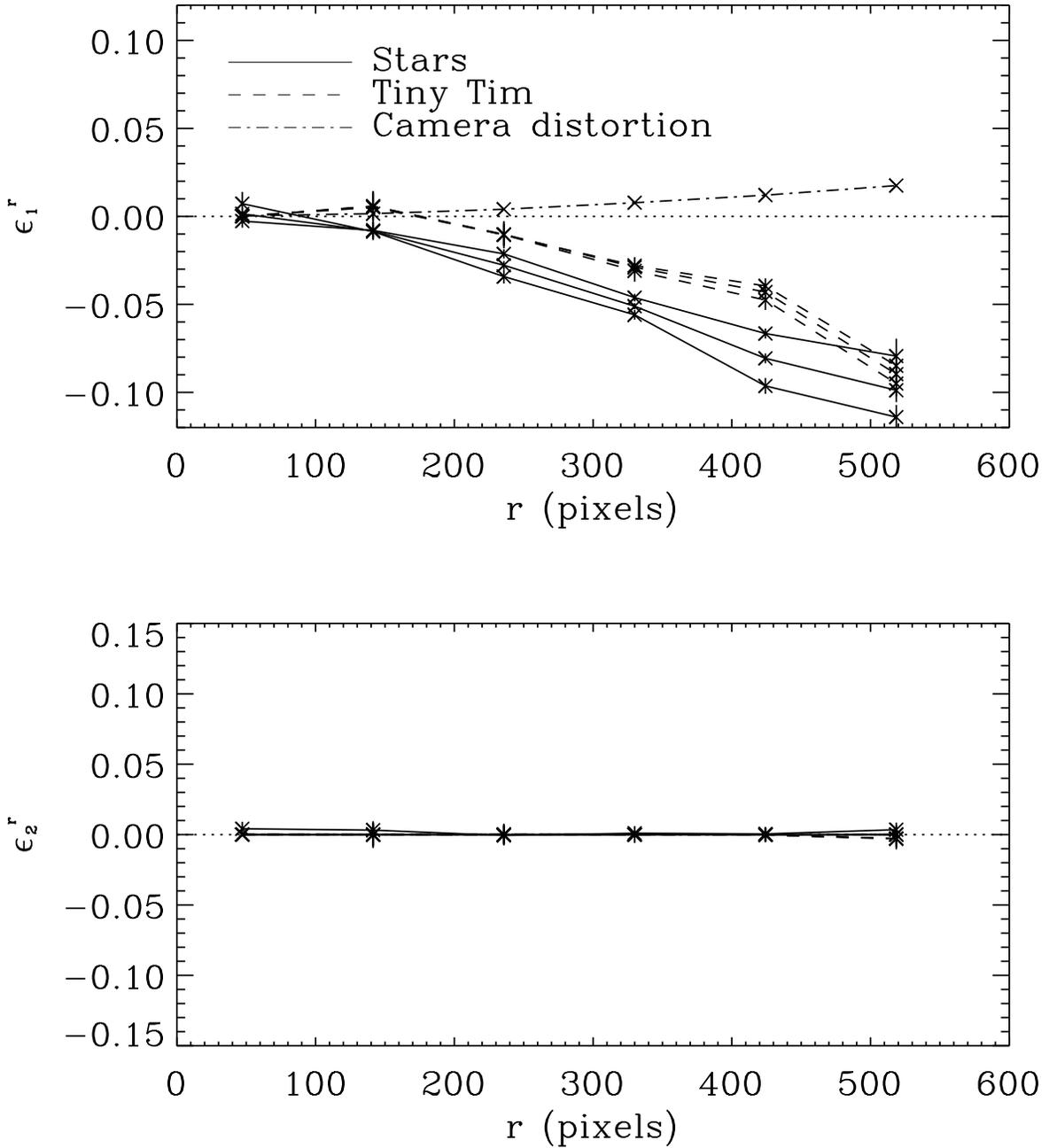}
\caption{Rotated ellipticity profile for the PSF and the camera
distortion, as a function of radius about the chip center. The solid
lines show the ellipticity as measured for M4, combined M4/NGC6572 ,
and NGC6572 stars, from top to bottom respectively. The dashed lines
show the PSF ellipticity predicted by Tiny Tim for focus values of
$+2.0$, $-1.5$, and $-5.0 \mu$m, from top to bottom respectively.  The
dot-dashed line shows the distortion ellipticity
$\epsilon_{i}^{\phi}$.}
\label{fig:inst}
\end{figure}

\begin{figure}
\plotone{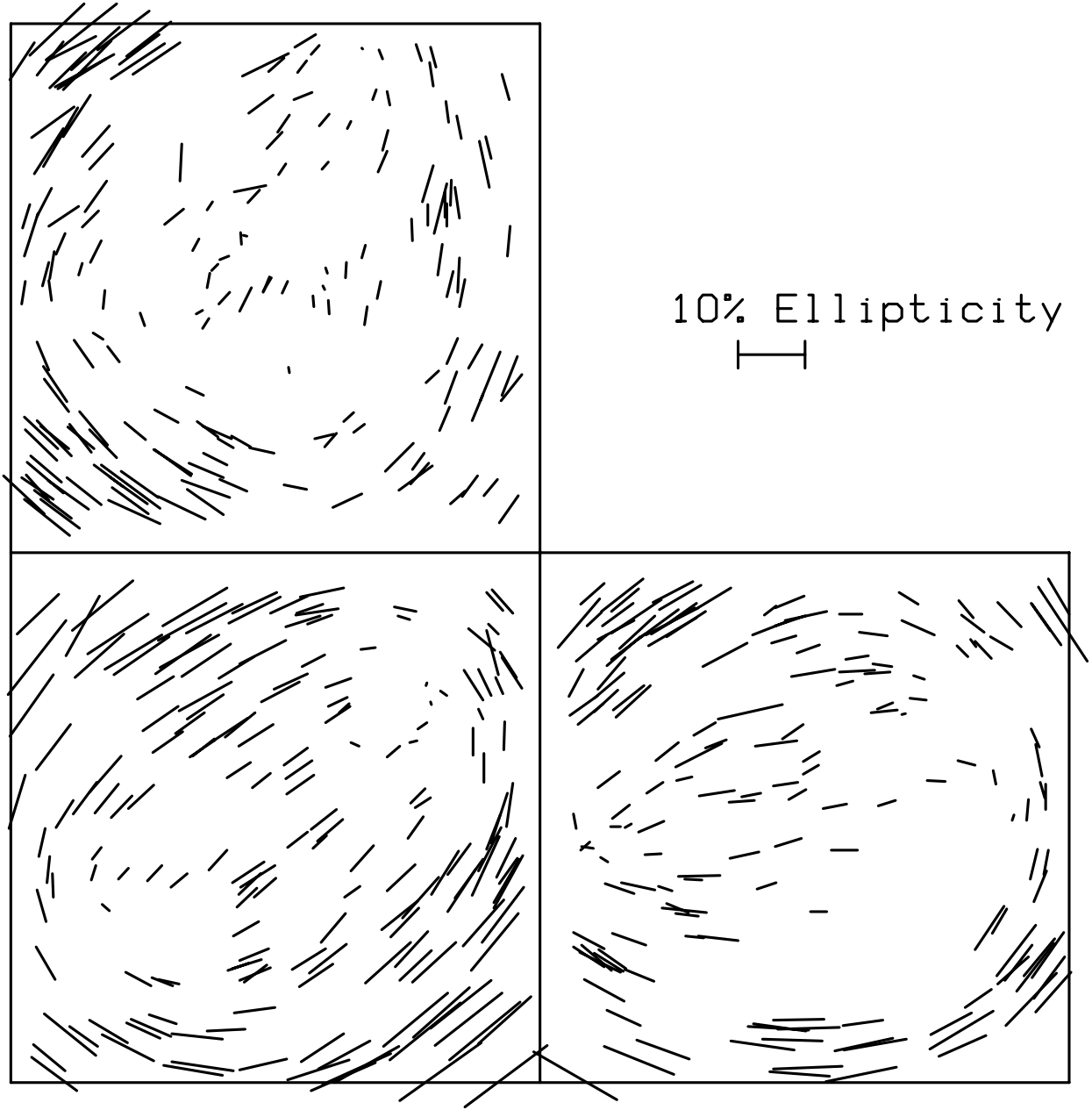}
\caption{Ellipticities of the stars in the globular cluster M4.
A Gaussian weight function of width $\omega_{*}=2$ pixels was used
to measure the stellar moments.}
\label{fig:m4ellipt}
\end{figure}

\begin{figure}
\plotone{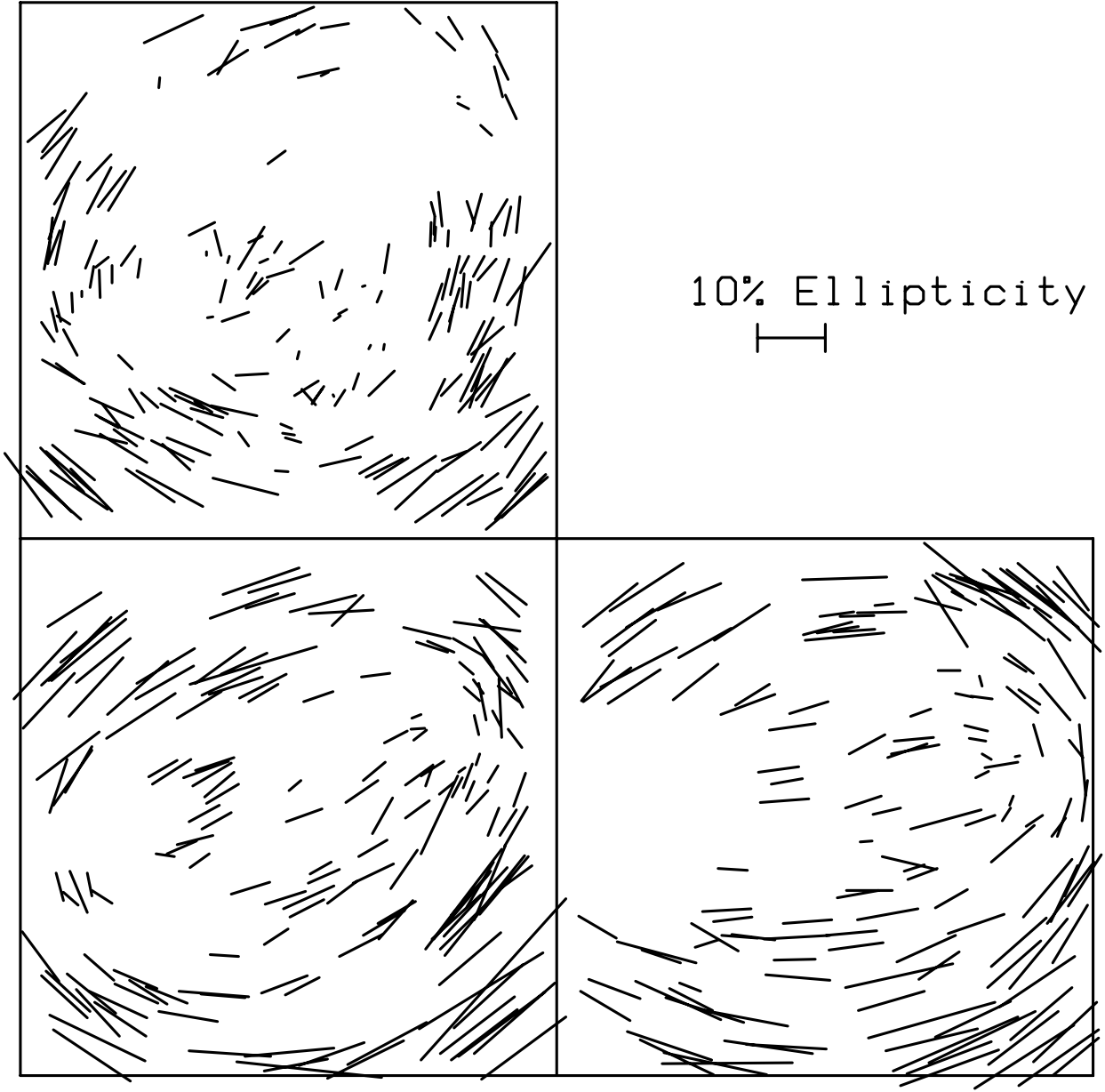}
\caption{Ellipticities of the stars in the globular cluster NGC6572.}
\label{fig:ngc6572}
\end{figure}

To study the variation of the PSF across the chip, we used Tiny Tim
to create a grid of PSFs. These PSFs were generated
for a telescope with a mirror at a focus position of $-1.5$ microns, the
value of the focus when the Survey Strip was taken (see
\S\ref{Strip}).  The resulting ellipticity pattern is shown in
Figure~\ref{fig:tt}. The corresponding ellipticity profile averaged
over all chips is shown as the central dashed line in
figure~\ref{fig:inst}. In these figures, we have added the effects of
the detector shear shown in figure \ref{fig:wf_shear}, so that a direct
comparison can be made between these simulations and the globular
cluster measurements which are presented below. The predicted PSF
anisotropy is quite large ($\langle \epsilon_{1}^{r} \rangle \simeq 0.1$ at
the edge) and varies significantly across each chip and from chip to
chip.

To test the effect of focus changes, we created Tiny Tim PSFs at
focus values of +2.0 microns and -5.0 microns, the full range of
telescope focus (see figure~\ref{fig:focus}). The ellipticity profiles
are shown are the upper and lower dashed curves in
figure~\ref{fig:inst}. The chip-averaged ellipticities for each focus
value are listed in table~\ref{tab:stats}. Extreme focus changes from
the central value of $-1.5$ microns produce variations of about 0.002
for $\langle \epsilon_{1}^{r} \rangle$ and about 0.006 for $\langle
\epsilon_{1} \rangle$ and $\langle \epsilon_{2} \rangle$, when
averaged over all chips.

\subsubsection{Globular Clusters}
For a more direct measurement of the PSF, we obtained archival images
of the globular cluster M4 observed by Richer et al.\
(1995).\markcite{ric97} We also obtained archival images of the
globular cluster NGC6572. This was the method of calibration used by
HFKS. It is fortuitous that the M4, NGC6572, and the Survey Strip
data all have focus values of approximately $-1.5$ microns (see
Figure~\ref{fig:focus}).

The M4 data consisted of 3 different fields in each
cluster. Each field had 8 different exposures:
four pairs of dithered pointings separated by very nearly 2 pixels.
Other dithers were separated by non-integer numbers of pixels.  
We shifted the dithers which are separated by integer numbers of
pixels and compared them pairwise to detect cosmic rays.  A pixel was
flagged as a cosmic ray if its flux was significantly above that of
the corresponding pixel in its pair image. We selected stars from the
M4 exposures by excluding saturated stars and those that are faint and
dominated by noise.  We also excluded stars confused by close
neighbors, and stars containing a cosmic ray flagged pixel within a 5
pixel radius. Our resulting sample consisted of about 160 stars per
chip. We then averaged the three moments $P_{kl}$ measured in each of
the eight exposures.  The analysis of the NGC6572 data was simplified
by the fact that we had only one field and the images were not
dithered.

\begin{figure}
\plotone{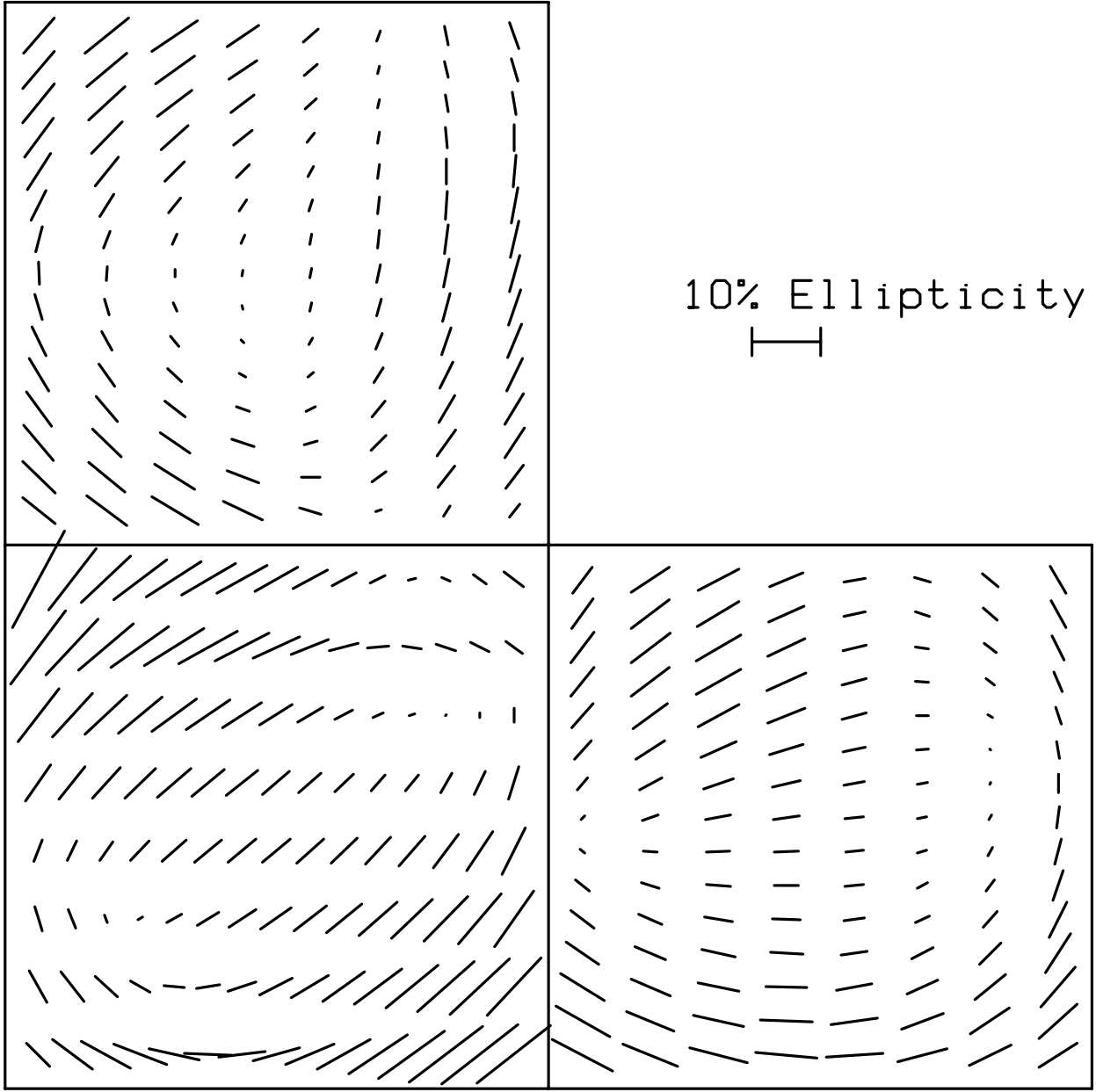}
\caption{PSF ellipticities derived from the combined M4/NGC6572 stars
after being corrected for shear and weighting and interpolated.}
\label{fig:m4sm}
\end{figure}

The ellipticity fields for M4 and NGC6572 are shown in
figures~\ref{fig:m4ellipt} and \ref{fig:ngc6572}, respectively. The
ellipticity profile for each globular cluster is shown as the upper
and lower solid line in figure~\ref{fig:inst}, respectively.  The
chip-averaged ellipticities for each globular cluster data set are
listed in table~\ref{tab:stats}. The M4 and NGC6572 ellipticity
patterns are qualitatively similar, and both show tangential
ellipticities of the order of 0.1 at the edge of all chips. The M4 and
NGC6572 average ellipticities $\langle \epsilon_{1}^{r} \rangle$ are
$-0.040$ and $-0.053$, respectively, and are both larger than that
predicted by Tiny Tim (about $-0.035$). The fact that the mean
tangential ellipticity for NGC6572 is larger than that for M4 by about
0.013, shows that the PSF varies substantially over periods as short
as a few days. This change is larger than that predicted by Tiny Tim
for extreme focus changes (see \S\ref{Tiny Tim}). Moreover, the
globular cluster ellipticities are qualitatively different from that
predicted by Tiny Tim. For example, figure~\ref{fig:m4sm} shows a
``teardrop'' shape to the M4 ellipticity pattern of chip 2, with a
cusp at the top of the chip. In contrast, the Tiny Tim ellipticity
pattern of the same chip, shown in figure~\ref{fig:tt} has a cusp on
the right-hand side.  Small changes in this cusp result by changing
the focus of the telescope in Tiny Tim, but it does not change the
qualitative pattern of the ellipticities.

Because of these shortcomings, we used the observed globular cluster
moments for the corrections, rather than the Tiny Tim predictions. To
reduce patchiness, we combined the M4 and NGC6572 stars, and corrected
the moments for the weight function
(Eqs.~[\ref{eq:pij_correction},\ref{eq:pijkl_correction}]) and for the
camera distortion
(Eqs~[\ref{eq:jij_dcorrection},\ref{eq:jijkl_dcorrection}] with
$\omega \rightarrow \infty$) to obtain the PSF moments $P_{ij}$ and
$P_{ijkl}$. We then fitted a fifth order polynomial in chip position
for each moment component and for each chip. Because the star counts
in the two globular cluster are close to one another (i.e. about 160
per chip), this amounts to giving equal weight to each of them. The
ellipticity profile for the combined globular clusters is shown as the
central solid line in figure~\ref{fig:inst}. Figure~\ref{fig:m4sm}
shows the resulting fitted ellipticity pattern of the PSF.

The mean PSF radius for the combined M4/NGC6572 was found to be
$g\simeq 0.89$ pixels (see Eq.~[\ref{eq:pij_decomp}]). This is much
smaller than the formal PSF radius that one would measure using a Tiny
Tim PSF with $\omega_{*} \rightarrow \infty$. As explained in
\S\ref{shortcomings}, the quadrupole moments indeed formally diverge
because of the extended wings of the HST PSF. Our value is however
commensurate with the PSF FWHM (0.9 pixels), and should thus be
considered as an effective PSF radius.
\label{psf_radius}

\subsection{Survey Strip}
\label{Strip}
The Survey Strip is a set of 28 contiguous pointings with the HST in
two colors, V (F606W) and I (F814W) (Groth et al.\
1995\markcite{gro95}; \markcite{rhot}Rhodes 1999). The images were taken
in March and April 1994 with the WFPC2. The Strip has about 10,000
galaxies down to $I\approx26$ and covers an area of about 108 square
arcmin over a $3'.5 \times 44'.0$ region. The Strip has already
proved a useful data set in exploring the distribution of luminous
matter through number counts and the 2-point angular correlation
function (Rhodes, Groth and Refregier 1998; Rhodes 1999).
\markcite{rha98} We are in the process of exploring the large scale
distribution of matter through weak lensing of the Strip galaxies
(\markcite{rhot99}Rhodes 1999; \markcite{rho99}Rhodes, Refregier, \&
Groth 1999).

Here, we use the Strip galaxies to test our shear measurement
method. For this purpose, we used the I images only.  The catalog of
galaxies was created using the Faint Object Classification and
Analysis System (FOCAS) within IRAF (Jarvis \& Tyson
1981)\markcite{jar81}. An object was considered as detected if two
contiguous pixels were more than $3\sigma$ above the sky background.
However, an object was  included in the final catalog only if both the
I and V images had a detection within an error box of approximately 5
pixels (0.5 arcsec). This resulted in a sample of 9448 galaxies
with $I<26$.

The moments of the galaxies were calculated as described above
(Eq.~[\ref{eq:jij_def}]). The variable weight function was
chosen to be $\omega=\rm max (2,\sqrt{ \frac{A}{\pi}})$, where $A$ is
the detection area calculated by FOCAS. The lower bound of
$\omega$ was chosen
to match the width $\omega_{*}$ used for stars.
\label{w_choice}

The importance of the camera distortion and of the PSF depends on the
radius $d$ of the galaxy (Eq.~[\ref{eq:d_def}]). We thus considered
subsamples of small and large galaxies, with $1.0 < d' <1.5$ pixels
and $d'>1.2$ pixels respectively. Here $d'$ is the root mean squared
observed radius (Eq.~[\ref{eq:d_obs}]) and $g\simeq 0.89$ pixels is
the PSF radius (see \S\ref{psf_radius}).  Because of the noise, a
fraction of the galaxies have unphysical ellipticity values
($\epsilon>1$). Because these outliers would dominate ellipticity
statistics, we only retained galaxies with $\epsilon<1$.  The mean
observed radius and reduction factor (Eq.~[\ref{eq:fd}]) are $\langle
d'\rangle \simeq 1.26$ pixels and $f_{d} \simeq 0.50$, and $\langle d'
\rangle \simeq 2.16$ pixels and $f_{d} \simeq 0.17$, for the small and
large galaxies, respectively.  Each subsample respectively comprises
41\% and 77\% of the total number of $I<26$ galaxies.

\begin{figure}
\plotone{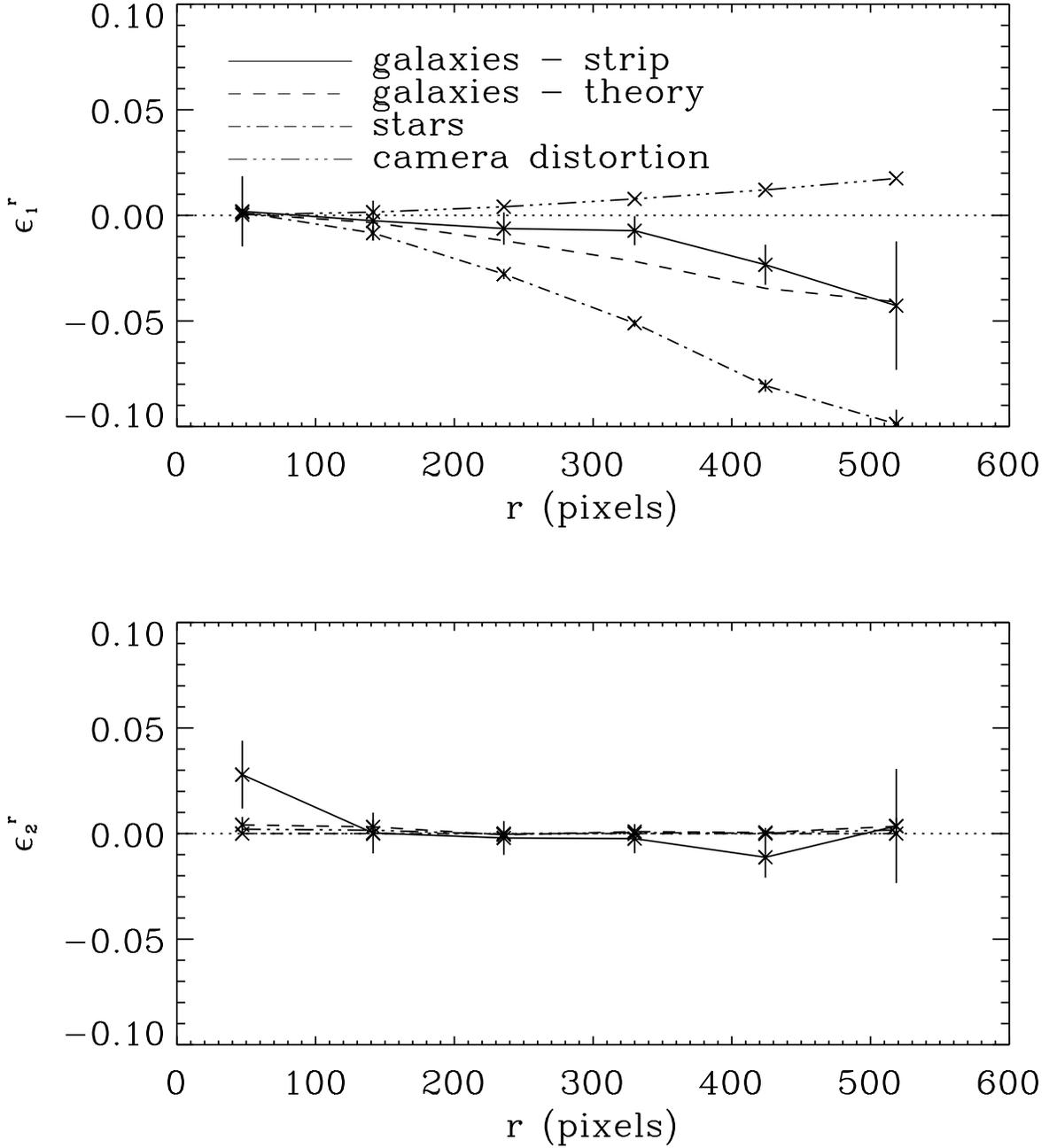}
\caption{Rotated ellipticity profile for small galaxies in the survey
strip. Galaxies were selected to have $I<26$, and radii $1.0<d'<1.5$
pixels. The PSF and the camera distortion ellipticity
$\epsilon_{i}^{\phi,r}$ is also shown for comparison. The profile for
the Strip galaxies expected in the simplified theoretical model is
also shown (see text).}
\label{fig:er_small}
\end{figure}

\begin{figure}
\plotone{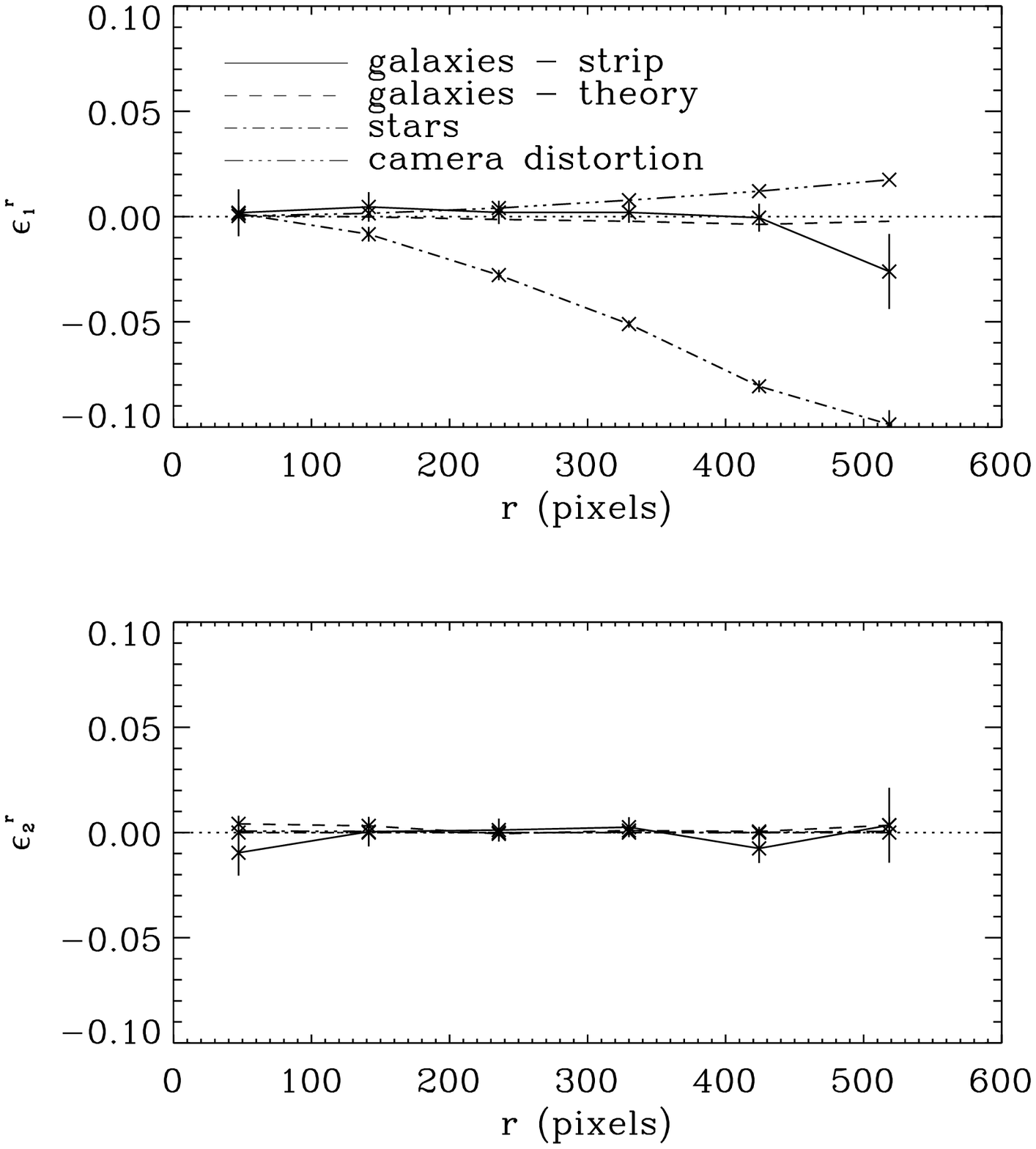}
\caption{Same as the previous figure, but for large galaxies ($d'>1.2$
pixels).}
\label{fig:er_large}
\end{figure}

The ellipticity profile for each subsample is shown as a solid  line in
figures~\ref{fig:er_small} and \ref{fig:er_large}. Also shown are the
asymptotic cases, namely the combined M4/NGC6572 stars ($d=0$), and
the camera distortion ellipticity ($d \rightarrow \infty$). The dashed
line shows the prediction of the simplified model described in
\S\ref{comb} (see Eq.~[\ref{eq:comb}]). The small galaxy sample shows
a tangential ellipticity which increases with radius from the chip
center, in agreement with the model prediction. The large galaxies, on
the other hand, do not display any significant tangential ellipticity,
again in agreement with the model. In both cases, $\langle
\epsilon_{2}^{r} \rangle$ is consistent with zero.

\begin{figure}
\plotone{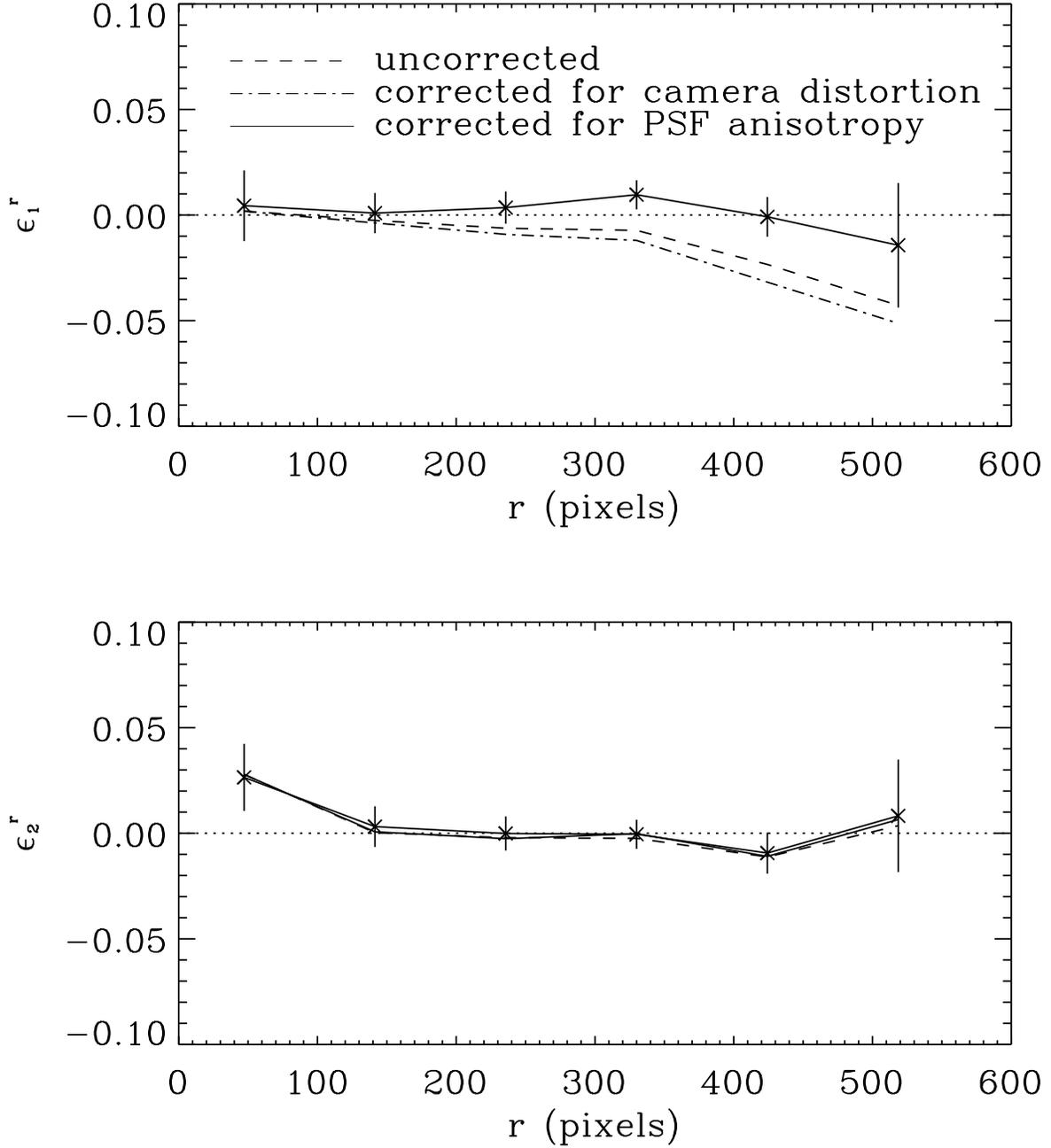}
\caption{Effect of the corrections for small galaxies ($1.0<d'<1.5$
pixels, $I<26$) in the survey strip. The profile for the galaxies is
shown at different stages of the correction algorithm.  The error bars
are similar for each of the three profiles, but, for clarity, were
only displayed for the last case.}
\label{fig:correction_small}
\end{figure}

\begin{figure}
\plotone{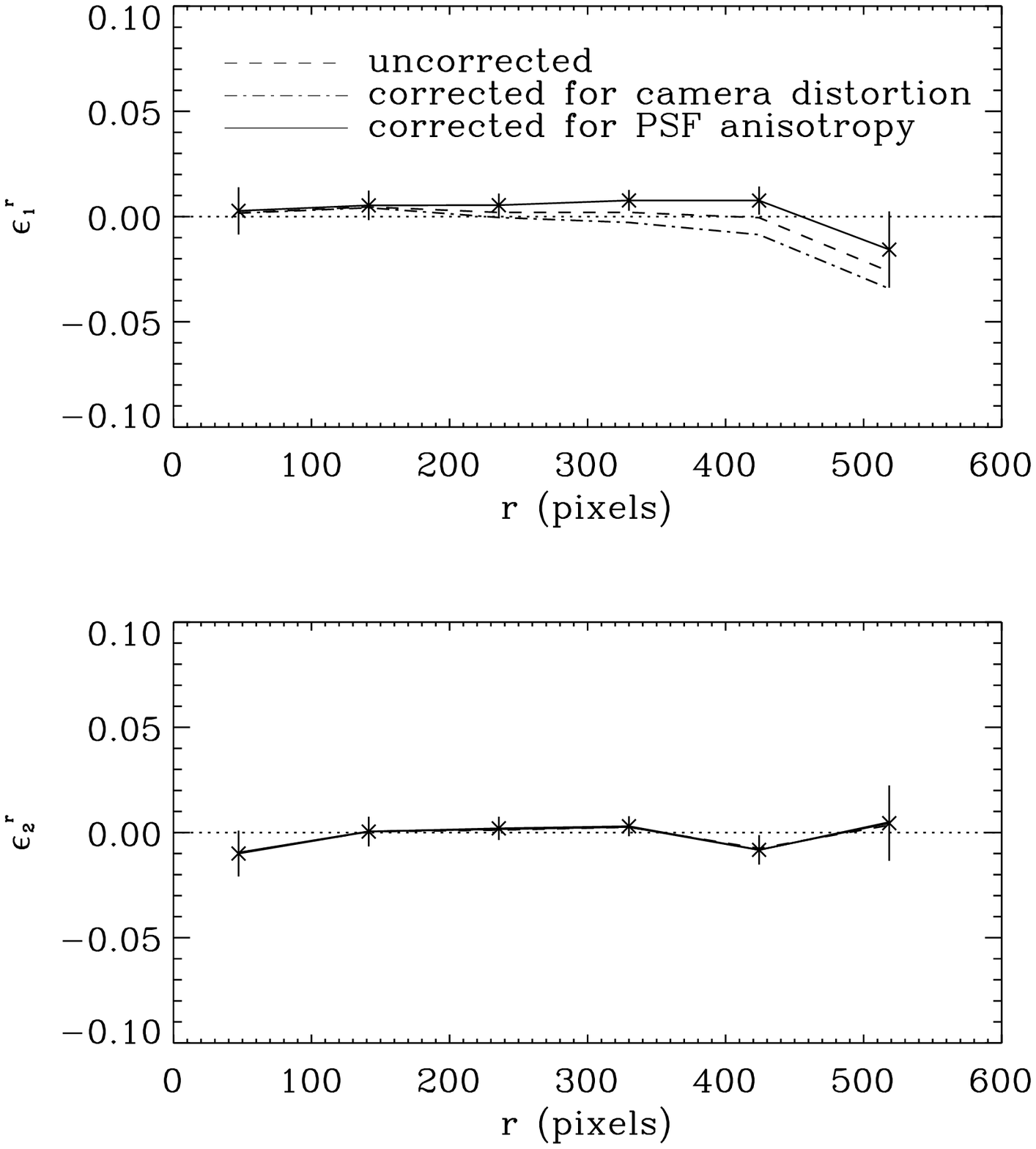}
\caption{Same as the previous figure, but for large galaxies
($d'>1.2$ pixels).}
\label{fig:correction_large}
\end{figure}

Using the camera distortion matrix derived in \S\ref{hst_distortion}
and the PSF moments derived in \S\ref{hst_psf}, we then corrected the
Strip galaxies. The ellipticity profiles at different stages of the
correction for the small and large galaxy subsamples are shown in
figures~\ref{fig:correction_small} and \ref{fig:correction_large}.
Since the camera distortion is radial, its correction increases the
tangential ellipticity of the galaxies. On the other hand, the
correction for the PSF, which is mainly tangential, reduces the
tangential component of the ellipticity. For both subsamples, the
resulting corrected profile is consistent with 0. As expected, the PSF
correction has a smaller effect on the large galaxies than on the
small galaxies.

\subsection{Error Budget}
\label{budget}
The effect of the corrections for the chip-averaged ellipticities are
summarized in table~\ref{tab:stats}. To ensure statistical
independence, we considered in this table a subsample of large
galaxies with $d'>1.5$ pixels (and $\epsilon<1$). This subsample
contains 51\% of the total number of galaxies with $I<26$,
corresponding to a surface density of $n\simeq 32$ arcmin$^{-2}$.
Their mean observed radius is $\langle d' \rangle \simeq 2.58$ pixels,
corresponding to a post-deconvolution reduction factor of $f_{d}^{\rm
deconv} \simeq 0.13$ (Eq.~[\ref{eq:fd_deconv}]). As a test of the
correction algorithm, we also corrected the globular cluster stars,
exactly as we corrected the galaxies.  (We have not corrected either
the stars or the small galaxies for the isotropic PSF, as this would
produce diverging ellipticities). The results for the corrected stars
are also shown in the table.

The self-corrected M4/NGC6572 stars have a residual ellipticity of
$\langle \epsilon_{1}^{r} \rangle \simeq .003$ and $\langle
\epsilon_{1} \rangle \simeq \langle \epsilon_{2} \rangle \simeq
.002$. This is a measure of the errors resulting from the
approximations in our correction method and from the fit to the
stellar moments. Another estimation of the errors from the method was
presented in \S\ref{validity}. By performing numerical simulations, we
showed that the residual ellipticity errors in the PSF correction was
0.004, after averaging over a galaxy ensemble with $d>1.5$ pixels.
Given the agreement of these two estimates, we take the error in the
correction method to be about 0.004.

As we noted in \S\ref{hst_psf}, the M4 and NGC6572 ellipticities
differ by about 0.01 for both rotated and absolute
ellipticities. After the anisotropic PSF correction, the small
galaxies have a residual ellipticity of $\langle \epsilon_{1}^{r}
\rangle \simeq 0.004$, $\langle \epsilon_{2} \rangle \simeq .011$ and
$\langle \epsilon_{1} \rangle \simeq .005$. Since the small galaxies
have a reduction factor of $f_{d} \simeq 0.50$, their residual
$\langle \epsilon_{1} \rangle$ ellipticity indicates that the PSF
variation is about 0.02, which is larger than that given by the
comparison of M4/NGC6572. We therefore take the variations of the
chip-average PSF ellipticity to be about 0.02.

How does this uncertainty affect the ellipticities of the large
galaxies? As noted above, the post-deconvolution reduction factor for
the large galaxies (with $d'>1.5$ pixels) is $f_{d}^{\rm deconv}
\simeq 0.13$.  Consequently, our uncertainty of 0.02 in the PSF
ellipticity produces an uncertainty in the ellipticities of the large
galaxies of only 0.003. This is of course a commendable consequence of
the small size of the WFPC2 PSF. Note also that, for the large
galaxies, the camera distortion correction is about the same size as
that for the PSF, so that the two corrections almost cancel each
other.  The changes in ellipticities produced by each of these
corrections is less than 0.005 for the large galaxies. The total
systematic uncertainty is a combination of the correction uncertainty
and of the PSF variability, and is thus about $0.004$. 

The statistical uncertainties are determined by the ellipticity
variance $\sigma_{\epsilon}^{2} \equiv \frac{1}{2} \langle
\epsilon^{2} \rangle = \frac{1}{2} \left( \langle \epsilon_{1}^{2}
\rangle + \langle \epsilon_{2}^{2} \rangle \right)$. Both noise in the
image and the intrinsic shapes of the galaxies contribute to this
dispersion. The {\it rms} ellipticity $\sigma_{\epsilon}$ is listed in the
last column of table~\ref{tab:stats}, at various stages of the
correction. For both the large and small galaxies, $\sigma_{\epsilon}$
does not increase when we correct for the camera distortion and for
the anisotropic PSF. This shows that our correction method does not
introduce any appreciable noise in the ellipticity measurements.  Not
surprisingly, $\sigma_{\epsilon}$ increases moderately when we correct
for the isotropic PSF. This is expected since this deconvolution
reduces the galaxy size, and thus reduces the denominator in the
definition of the ellipticity. For the large galaxies, the {\it rms}
ellipticity after all corrections is about $\sigma_{\epsilon} \simeq
0.31$. Since the mean number of such galaxies per chip is about
$N_{g}\simeq 57$, the $1\sigma$ sensitivity to detect the shear in a
chip is $\sigma_{\epsilon}/\sqrt(N_{g}) \simeq .04$. This is close to
the expected {\it rms} shear expected from weak lensing by large-scale
structure in 1.3 arcmin cells, for cluster normalized CDM models and
for a source redshift $z=1$ (\markcite{jain97} Jain \& Seljak 1997).
This shows that the expected signal-to-noise ratio of a single WFPC2
chip is about 1.

\section{Conclusions}
\label{conclusions}
We have revisited the KSB method to measure the weak lensing shear
from the shapes of galaxies. In our method, the corrections for the
camera distortion and PSF convolution are performed using moments
rather than ellipticities. Using a gaussian weight function, we
derived explicit expressions for the corrections, which involve only
second and fourth order moments. We clarified the convergence of some
of the approximations made by KSB, and showed how the weight function
for stars can be chosen to be different from that for galaxies.  We
also showed how the isotropic part of the PSF can be assumed to be a
gaussian, to the required level of precision, and can thus be
corrected for exactly. We derived the explicit relation between the
shear and the ellipticities by decomposing moments into tensors with
definite rotational properties.

We addressed the recently exposed shortcomings of the KSB method
(\markcite{kai99}Kaiser 1999; \markcite{kui99}Kuijken 1999).  Our
method, as well as the KSB method, has formal problems arising from
the fact that PSFs encountered in practice are not sufficiently
compact. We used numerical simulations to assess the importance of
these problems in the analysis of WFPC2 images. We found that the
ellipticity error produced by the PSF correction can be of several
percent for an individual galaxy. However, it is only about 0.004 when
averaged over a galaxy ensemble with $d>1.5$ pixels, provided moderate
weight function widths are used.

We studied systematic effects arising in WFPC2 images. From globular
cluster observations, we confirm the results of HFKS, who found
$\sim$10\% PSF ellipticities at the edge of each chip. We find however
that the camera distortion is radial, rather than tangential. It
produces ellipticities of the order of 0.7\%. We further find that the
PSF ellipticity varies by as much as 2\% over time.

We applied our correction method to the HST Survey Strip.  We showed
that the different stages of our correction do not introduce any
appreciable noise. We studied the dependence of galaxy ellipticities
on the galaxy size. Small galaxies are more sensitive to the PSF and
also indicate that the PSF varies with time. For large galaxies
(observed radii $d'>1.5$ pixels), the total systematic uncertainty is
about 0.4\%, and results from a nearly equal contribution from the
correction uncertainty and from the PSF variability. The statistical
$1\sigma$ uncertainty in measuring the shear in a single WFPC2
$1.3\times1.3$ square arcmin chip is about 4\%, for this subsample of
galaxies. This provides good prospects for detecting a cosmic shear
signal with the strip and other deep HST surveys. In
\markcite{rho99}Rhodes et al.  (1999) and \markcite{rhot99}Rhodes
(1999), we will describe our search for such a signal.

\acknowledgments We thank Richard Ellis, Peter Schneider, Stella
Seitz, Henk Hoekstra, Christophe Alard, David Bacon and Meghan Gray
for useful discussions. We are grateful to Nick Kaiser, the referee,
for insightful comments and criticisms. AR was supported by the NASA
MAP/MIDEX program and by the NASA ATP grant NAG5-7154.  EG and JR were
supported by NASA Grant NAG5-6279, and would like to thank the WFPC1
IDT for cooperation on this project.

\newpage

\begin{deluxetable}{lcrrrrr}
\footnotesize
\tablecaption{Effect of Corrections on Chip-Averaged Ellipticities
\label{tab:stats}}
\tablewidth{0pt}
\tablehead{
\colhead{Sample} &
\colhead{Corrections\tablenotemark{a}} &
\colhead{$\langle \epsilon_{1}^{r} \rangle$} &
\colhead{$\langle \epsilon_{2}^{r} \rangle$} &
\colhead{$\langle \epsilon_{1} \rangle$} &
\colhead{$\langle \epsilon_{2} \rangle$} &
\colhead{$\sigma_{\epsilon}$\tablenotemark{b}}
}
\startdata
Camera distortion & &
  $ 0.007 \pm 0.001$ & $-0.000 \pm 0.001$ & $ 0.000 \pm 0.001$ &
  $-0.000 \pm 0.001$ &  \\
M4+NGC6572 &  & 
  $-0.047 \pm 0.001$ & $ 0.001 \pm 0.001$ & $-0.014 \pm 0.001$ & $-0.020
  \pm 0.002$ & \\
                        & d,a &
  $-0.003 \pm 0.001$ & $ 0.000 \pm 0.001$ & $-0.002 \pm 0.001$ & $ 0.002
  \pm 0.001$ & \\
M4               &  &
  $-0.040 \pm 0.002$ & $-0.001 \pm 0.001$ & $-0.010 \pm 0.001$ &
  $-0.020 \pm 0.002$ & \\
                 & d,a &
  $ 0.004 \pm 0.001$ & $ 0.000 \pm 0.001$ & $ 0.003 \pm 0.001$ & $
  0.003 \pm 0.001$ & \\
NGC6572   &  &
  $-0.053 \pm 0.002$ & $ 0.003 \pm 0.002$ & $-0.019 \pm 0.002$ &
  $-0.020 \pm 0.003$ & \\
          & d,a &
  $-0.009 \pm 0.001$ & $-0.000 \pm 0.001$ & $-0.007 \pm 0.001$ & $
  0.001 \pm 0.001$ & \\
Tiny Tim\tablenotemark{c} ($-1.5\mu$) &  &
  $-0.035 \pm 0.003$ & $-0.000 \pm 0.003$ & $-0.032 \pm 0.002$ &
  $-0.031 \pm 0.003$ & \\
~~~~~~~~~~~~~~($+2.0\mu$)  &  &
  $-0.033 \pm 0.003$ & $-0.001 \pm 0.002$ & $-0.025 \pm 0.001$ &
  $-0.023 \pm 0.003$ & \\
~~~~~~~~~~~~~~($-5.0\mu$) &  &
  $-0.038 \pm 0.004$ & $-0.000 \pm 0.003$ & $-0.038 \pm 0.002$ &
  $-0.037 \pm 0.004$ & \\
Small galaxies\tablenotemark{d} &  &
  $-0.009 \pm 0.004$ & $-0.002 \pm 0.004$ & $ 0.005 \pm 0.004$ &
  $-0.005 \pm 0.004$ & 0.24 \\
  & d &
  $-0.013 \pm 0.004$ & $-0.001 \pm 0.004$ & $ 0.004 \pm 0.004$ &
  $-0.004 \pm 0.004$ & 0.24 \\
  & d,a &
  $ 0.004 \pm 0.004$ & $ 0.000 \pm 0.004$ & $ 0.011 \pm 0.004$ &
  $0.005 \pm 0.004$ & 0.24 \\
Large galaxies\tablenotemark{d} & &
  $ 0.004 \pm 0.003$ & $-0.000 \pm 0.003$ & $ 0.006 \pm 0.003$ &
  $-0.005 \pm 0.003$ & 0.24 \\
  & d &
  $ 0.001 \pm 0.003$ & $-0.000 \pm 0.003$ & $ 0.007 \pm 0.003$ &
  $-0.006 \pm 0.003$ & 0.24 \\
  & d,a &
  $ 0.005 \pm 0.003$ & $ 0.000 \pm 0.003$ & $ 0.009 \pm 0.003$ &
  $-0.003 \pm 0.003$ & 0.24 \\
  & d,a,i &
  $ 0.008 \pm 0.004$ & $-0.000 \pm 0.004$ & $ 0.009 \pm 0.004$ &
  $-0.001 \pm 0.004$ & 0.31 \\
\enddata
\tablenotetext{a}{Corrections for: d: camera distortion,
  a: PSF anisotropy, i: isotropic PSF. The PSF is derived from
  the combined M4+NGC6572 stars.} 
\tablenotetext{b}{$\sigma_{\epsilon}^{2} \equiv
 \frac{1}{2} \langle \epsilon^{2} \rangle = \frac{1}{2}
 \left( \langle \epsilon_{1}^{2} \rangle + \langle \epsilon_{2}^{2} \rangle
  \right)$}
\tablenotetext{c}{PSF predicted by Tiny Tim for several
  focus values}
\tablenotetext{d}{Galaxies with magnitudes $I<26$, ellipticities
  $\epsilon<1$,
 and observed radii $1.0<d'<1.5$ and $d'>1.5$ pixels, for the
 small and large sample,
respectively. }
\end{deluxetable}

\end{document}